\documentclass[12pt, draftclsnofoot, onecolumn]{IEEEtran}
\usepackage{}
\usepackage{amsmath}
\usepackage{graphicx}
\usepackage{amssymb}
\usepackage{amsmath,amsthm}
\usepackage{float}
\usepackage{enumerate}
\usepackage{epstopdf}
\usepackage{dblfloatfix}
\usepackage{caption}
\usepackage{amsfonts}
\usepackage{fancyhdr}
\usepackage{bbm}
\usepackage{cases}
\usepackage{cite}
\usepackage{breqn}
\usepackage{multirow}
\usepackage{mathrsfs}
\usepackage{algpascal}
\usepackage{algc}
\usepackage{algorithmicx}
\usepackage[ruled]{algorithm}
\usepackage{algpseudocode}
\usepackage{graphics}
\usepackage{epsfig}
\usepackage{mathrsfs}
\usepackage{dsfont}
\usepackage{tabularx}
\usepackage{color}
\usepackage{booktabs}
\newtheoremstyle{theorem}
  {\topsep}
  {\topsep}
  {\itshape}
  {}
  {\itshape}
  {:}
  {.5em}
  {\thmname{#1}\thmnumber{ #2}\thmnote{ (#3)}}
\theoremstyle{theorem}
\newtheorem{theorem}{Theorem}

\newtheorem{proposition}{Proposition}

\newtheorem{definition}{Definition}

\newcommand{\tocaption}{%
\setlength{\abovecaptionskip}{0pt}%
\setlength{\belowcaptionskip}{10pt}%
\caption}

\floatname{algorithm}{Algorithm}

\makeatletter
\newcommand{\multiline}[1]{%
  \begin{tabularx}{\dimexpr\linewidth-\ALG@thistlm}[t]{@{}X@{}}
    #1
  \end{tabularx}
}
\makeatother
\algdef{SE}[DOWHILE]{Do}{doWhile}{\algorithmicdo}[1]{\algorithmicwhile\ #1}
\DeclareMathOperator*{\argmin}{argmin}
\IEEEoverridecommandlockouts
\begin{document}
\title{Mobility-Aware Routing and Caching in Small Cell Networks using Federated Learning}
\author{Yuwen Cao\textsuperscript{},~\IEEEmembership{Student Member, IEEE}, Setareh Maghsudi\textsuperscript{},~\IEEEmembership{} and
Tomoaki Ohtsuki\textsuperscript{},~\IEEEmembership{Senior Member, IEEE}
\thanks{This paper was presented in part at the IEEE Conference on Communications (ICC) 2021 \cite{9500804}. Y. Cao and T. Ohtsuki are with the Graduate School of Science and Technology, Keio University, Yokohama 223-8522, Japan (e-mail:ywcao@ohtsuki.ics.keio.ac.jp; ohtsuki@ics.keio.ac.jp). S. Maghsudi is with the Department of Computer Science, University of T\"{u}bingen, 72074 T\"{u}bingen- Germany, and with the Fraunhofer Heinrich-Hertz-Institute, 10587 Berlin- Germany (email:setareh.maghsudi@uni-tuebingen.de). The work of Y. Cao and T. Ohtsuki was supported by JSPS KAKENHI Grant Number JP20J12528. The work of S. Maghsudi was supported by Grant 16KIS1165 from the German Federal Ministry of Education and Research.}
}
{}
\maketitle
%
\begin{abstract}
We consider a service cost minimization problem for resource-constrained small-cell networks with caching, where the challenge mainly stems from (i) the insufficient backhaul capacity and limited network bandwidth and (ii) the limited storing capacity of small-cell base stations (SBSs). Besides, the optimization problem is NP-hard since both the users' mobility patterns and content preferences are unknown. In this paper, we develop a novel mobility-aware joint routing and caching strategy to address the challenges. The designed framework divides the entire geographical area into small sections containing one SBS and several MUs. Based on the concept of one-stop-shop (OSS), we propose a federated routing and popularity learning (FRPL) approach in which the SBSs cooperatively learn the routing and preference of their respective MUs and make a caching decision. The FRPL method completes multiple tasks in one shot, thus reducing the average processing time per global aggregation of learning.
By exploiting the outcomes of FRPL together with the estimated service edge of SBSs,
the proposed cache placement solution greedily approximates the minimizer of the challenging service cost optimization problem.
Theoretical and numerical analyses show the effectiveness of our proposed approaches.
\end{abstract}
{\it Index Terms}:
Caching, federated learning, routing, mobility patterns, small cell network, one-stop-shop, multiple tasks.
\section{Introduction}
\label{sec:Intr}
Wireless caching is a promising concept to reduce the peak traffic and the backhaul load, particularly for video content delivery \cite{6708492,6600983,9080557}. The caching concept originates from the \textit{content reuse property} of video streaming, i.e., users are likely to request the same video content. Thus saving popular content at the local small-cell base stations (SBSs) during the off-peak time or pushing them at user devices directly through broadcasting improves the network's throughput and coverage performance. Besides, it enhances the users' quality of experience (QoE) \cite{7807344,li2016trend}.

Realizing the potential of wireless caching necessitates optimizing several parameters. Under lack of sufficient prior knowledge, one can take advantage of a recently-emerged concept, namely \textit{federated learning} (FL). FL enables several partners to jointly learn the parameters of a specific model in a distributed manner, i.e., without requiring any data exchange \cite{article201602}. Thus, using FL strongly reduces the amount of data uploaded via the wireless uplink channel. Besides, FL maintains the benefits of cognitive- and swift reaction to the mobility and other characteristics of cellular networks, as well as preserving personal data privacy \cite{lim2019federated}.
Besides, the FL concept is a widespread tool for wireless edge network optimization \cite{lim2019federated,yang2019scheduling,zhu2018low,li2020federated,9292468}, due to the following three reasons: (i) The ever-increasing number of devices in the internet of things (IoT) generate an enormous amount of data at the network edge. That entails an enabling technology and efficient approaches for \textit{sensitive data management and utilization} in wireless edge network; (ii) With the computation- and storage constraints of increasingly complex edge networks, conventional network optimization approaches built on static models perform poorly in modeling dynamic networks;
(iii) Conventional machine learning-based approaches can optimize decision-making through interactions with the dynamic environment. However, such methods require user data as input, which might be sensitive or inaccessible in nature due to regulatory constraints. Thus FL is an enabling technology for resource-constrained network optimization (concerning computation and storage resources), and for bringing intelligence to the wireless edge network.

Reference \cite{7448844} formulates a joint link scheduling and power allocation problem for (device-to-device) D2D-assisted wireless caching network. The authors then decompose the original problem into two subproblems and alternately solve subproblems by using the convex optimization method \cite{boyd2004convex}. In \cite{8393473}, \textit{Want et al.} investigate the NP-hard cache placement problem for D2D-assisted network. The authors propose a dynamic programming algorithm that greedily assigns the contents to the empty caches. In \cite{7968495}, the authors investigate an energy-efficient transmission mode selection problem for D2D-assisted small-cell networks. By relaxing the hard problem and using the bandit theory \cite{maghsudi2014transmission}, they achieve the maximum network utility. Furthermore, reference \cite{7438743} studies a joint transmission and caching policy optimization problem for D2D-assisted small-cell networks, given that the users' demands are known a prior.
Wireless caching in small-cell networks is also studied in \cite{7775114,muller2016smart,7422747,maghsudi2019bandit}. Reference \cite{7775114} develops a context-aware proactive caching approach for small-cell networks. Such an approach and its variant in \cite{muller2016smart} learn the context-specific content popularity online. In \cite{7422747}, the authors introduce a learning-based approach to caching in heterogeneous small-cell networks.
Besides, \cite{maghsudi2019bandit} investigates a joint femto-caching and power control problem in small-cell networks without knowing any prior information on file popularity.

References \cite{7370924,6763007,7417276} study caching based on network coding (\textit{coded-multicast}). \cite{7370924} investigates the caching policy optimization based on multicast transmissions. The results reveal that properly combining caching and multicast transmission reduces energy costs. By jointly optimizing the cache placement and multicast transmission, \cite{6763007} proposes a novel coded caching approach that can guarantee a \textit{global caching gain}. In \cite{7417276}, the authors develop an optimal femto-cache placement strategy that maximizes the number of network coded file downloads from femto-caches, thereby significantly reducing the macro-cell base station (MBS) bandwidth usage.
%
\begin{table*}
\captionsetup{format=plain, singlelinecheck=off, labelsep=newline, font={small,sc}}
\captionsetup{justification=centering}
\tocaption{\newline \footnotesize{The State-Of-The Art Research of Caching in Small-Cell Networks}}
  \centering
  \scriptsize
  \begin{tabular}{|c|c|c|c|c|}
    \hline
    \textbf{Ref.} & \textbf{Optimization Problem}  & \textbf{Prior Information} & \textbf{Resource Constrained} & \textbf{Merit}
       \\
       \hline
       \hline
        [15] & {Transmission, Caching policy} & {Popularity} & {None} & {Low energy and economical cost} \\
        \hline
       [16] & {Cache placement}  & {User context} & {Yes} & {High cache hit ratio}\\
       \hline
       [18] & {Cache placement}  & {User demand} & {None} & {A lower bound on the training time} \\
       \hline
       [22] & {Transmission, Caching policy}  & {Popularity} & {None} & {Low energy cost}\\
       \hline
       [25] & {Cache placement} & {Popularity}  & {None} & {Low computation complexity} \\
       \hline
       [26] & {Energy efficiency, Cache placement}  & {Popularity, Perfect CSI} & {None} & {High cache hit ratio} \\
       \hline
       [27] & {Caching utility}  & {Popularity, Mobility data} & {None} & {High caching utility}\\
              \hline
       \text{Our work} & {Routing, Cache placement}  &  {None} & {Yes}  & {See \textbf{Section \ref{subSec:contri}}} \\
       \hline
  \end{tabular}
  \label{Tb:Summary}
\end{table*}

Mobility-aware caching exploits users' mobility statistics for allocating caching resources \cite{7932468,8067654,7037523,9042255}. Reference \cite{7932468}, e.g., proposes a mobility-aware cache placement strategy that maximizes the data offloading ratio in D2D-assisted networks, given precise prior information of file popularity. Similarly, \cite{8067654} assumes that each SBS has the global channel state information (CSI) and file popularity. Based on this assumption, the authors develop a green mobility-aware caching model to jointly optimize the cache placement and power allocation among SBSs and mobile devices. Besides, \cite{7037523} explores the mobility-aware content caching problem for small-cell networks to maximize the caching gain given mobile users' (MUs') preferences and MUs' mobility patterns. Using prior knowledge about the MUs' locations at a central processor, \cite{9042255} proposes a content cooperative caching approach that maximizes the sum mean opinion score (MOS) of MUs. \textbf{Table \ref{Tb:Summary}} summarizes the state-of-the-art research in caching for resource-constrained small-cell network.
\subsection{Summary of Contributions}
\label{subSec:contri}
In the majority of references mentioned above, the popularity profile of content files is either known a prior or modeled as a Zipf-like distribution \cite{breslau1999web}\cite{hefeeda2008traffic}. However, such an assumption is unrealistic in practice, particularly for the dense small-cell networks: (i) The file popularity is often non-stationary due to the dynamic nature of contents; (ii) Perfect knowledge of file popularity is technically difficult to guarantee, especially for intensive instantaneous demands; (iii) File popularity is an outcome of the decisions of a dense population, and very costly to predict \cite{9080557}. In such a situation, developing a low-cost and effective learning-based approach for estimating the files' popularity becomes imperative. Besides, to our best knowledge, no research has considered joint routing and cache placement optimization for resource-constrained small-cell networks so far.

Against this background, we focus on developing mobility-aware routing and caching strategies for dynamic small-cell networks under uncertainty. To jointly optimize the routing and cache placement, we first formulate {\it the network cost} minimization problem, which is an NP-hard integer programming (NIP) problem. We propose a federated routing and popularity learning (FRPL) approach by which the SBSs learn the popularity of files as a function of the non-stationary pedestrian- and request densities. To avoid unnecessary data retransmission via the backhaul, we develop a novel content transmission protocol that improves the cache-efficiency (CE) performance of SBSs. Besides, we optimize the cache placement using an algorithm that greedily approximates the minimizer of the NIP problem. The contributions of this paper are as follows:
\begin{itemize}
\item Motivated by the notion of the one-stop-shop (OSS), we propose an FRPL approach that enables SBSs to complete multiple tasks simultaneously, thus reducing the processing time. That is crucial as the time slot within one global aggregation is limited.
\item The proposed FRPL method jointly learns the routing and content popularity considering the influence of both the users' mobility and the SBSs' geographical location. Therefore, it is robust against even in a highly dynamic and non-stationary network.
\item We design a cache placement approach that exploits the outcome of FRPL method combined with a novel transmission protocol to minimize the network's content delivery cost by optimizing the cached content and delivery strategy.
\item Numerical results show that our cache placement policy guarantees a higher cache hit rate for SBSs compared to the existing schemes, although the MUs' demand density is a priori unknown. Besides, the proposed greedy cache placement approach reduces the network cost compared to the state-of-the-art research. We also show the effectiveness of the proposed FRPL approach.
\end{itemize}

In \textbf{Section \ref{sec:Sys}}, we present the network model, FL model, and content request model. We then formulate the associated resource-constrained network optimization problem. In \textbf{Section \ref{sec:FedRout}}, we propose an FRPL approach for multi-task learning. The analysis of the proposed model and solution appear in \textbf{Section \ref{Sec:CachePlacement}}. \textbf{Section \ref{sec:simulation}} includes the numerical analysis. \textbf{Section \ref{sec:Conclusion}} concludes the paper.
%
\begin{table*}
\captionsetup{format=plain, singlelinecheck=off, labelsep=newline, font={small,sc}}
\captionsetup{justification=centering}
\tocaption{\newline \footnotesize{List of Most Important Variables}}
  \centering
  \scriptsize
  \begin{tabular}{|c|c|}
    \hline
    \textbf{Symbols} & \textbf{Definitions/Explanations}
       \\
       \hline
       \hline
      $Q_{i}$ &  {The number of the samples collected by a participant $i$.} \\
       \hline
      $\boldsymbol{\rho} _{i}$ &  {The local FL model hyperparameter.} \\
       \hline
      $\boldsymbol{\beta} =  \frac{1}{Q} \sum\nolimits_{i=1}^{{I}} Q_{i}  \boldsymbol{\rho} _{i}$ & {The update of the global FL model.}\\
       \hline
      $p_{f}$ & {The probability that a request for file $f \in \{1, \cdots, M\}$ at time $t$, and $p_{f} \in [0,1]$.} \\
       \hline
       $c^{}_{k,f}$ & {The binary caching strategy of file $f$ at SBS $k \in\{1,\cdots,K\}$ at time $t$, and $c^{}_{k,f} \in \{0,1\}$.} \\
      \hline
      $\psi_{k,t}$ & {The pedestrian density at time $t$ at the site of SBS $k$.} \\
       \hline
      $\lambda_{f,t}$ & {Expected request density of a specific file $f$ at time $t$.} \\
       \hline
       $C_{k}^{S}$ & {Cache capacity of SBS $k$.} \\
       \hline
       $\alpha_{C_f}$ & {Cost incurred by caching file $f$ at each SBS.}\\
       \hline
       $\beta_{M}$; $\beta_{M_k}$; $\beta_{S_k}$ & Costs incurred by retrieving file $f$ from backhaul to MBS, from MBS to SBS $k$, or directly from SBS $k$. \\
       \hline
       $d^{}_{k,f}$ & {Normalized costs incurred by retrieving file $f$ from MBS to SBS $k$.} \\
       \hline
       $L_{1}({\boldsymbol{\rho}_{1,i}})$; $L_{2}({\boldsymbol{\rho}_{2,i}})$ & {Loss function for training local FL model hyperparameters ${\boldsymbol{\rho}_{1,i}}$ (${\boldsymbol{\rho}_{2,i}}$) w.r.t. the pedestrian (request)-density prediction task.}\\
       \hline
       $(x_{k}(t),y_{k}(t))$ & {The final location of the centroid of cluster $k$.} \\
       \hline
        $N_{k,t}^{+}(j)$; $N_{k,t}^{-}(j)$ & {Statistical function of pedestrian density for those who might come (leave) cell $k$ from (to) neighboring cell $j$.} \\
      \hline
       $\nabla  f(\boldsymbol{\beta}_{q,t},\mathbf{x}_{q,i,j},{y}_{q,i,j})$ & {The gradient function of $f(\boldsymbol{\beta}_{q,t},\mathbf{x}_{q,i,j},{y}_{q,i,j})$ w.r.t. the global FL model $\boldsymbol{\beta}_{q,t}$. }\\
       \hline
       $\nabla F(\boldsymbol{\beta}_{q,t},\mathbf{x}_{q,i,j},{y}_{q,i,j})$ & {The gradient function of $F(\boldsymbol{\beta}_{q,t},\mathbf{x}_{q,i,j},{y}_{q,i,j})$ w.r.t. the global FL model $\boldsymbol{\beta}_{q,t}$. }\\
       \hline
        $\mathcal{L}_{k,t}$ & {The serving edge consisting of a set of MUs of which SBS $k$ will serve at time $t$.} \\
       \hline
  \end{tabular}
  \label{Tb:Symbols}
\end{table*}

\textit{Notations}: Bold lower case letters represent vectors, whereas bold upper case letters denote matrices. $\left[ {\cdot} \right]^{T}$, $\left\| {\cdot} \right\|_{0}$, and $\left\| {\cdot} \right\|$ denote transpose, $l_{0}$-norm, and $l_{2}$-norm operations, respectively. Besides, $\mathbb{E}[\cdot]$ is used to denote the statistical expectation operation. $\nabla$ represents the gradient of a function.
\textbf{Table \ref{Tb:Symbols}} summarizes the frequently-used variables in the order of appearance in the paper.
%
\section{System Model and Problem Formulation}
\label{sec:Sys}
%
\subsection{Network Model}
We consider a network consisting of $K$ SBSs and one MBS. The network serves $U$ MUs simultaneously. Due to the dense deployment, the coverage areas of the SBSs and the MBS overlap; Hence, at each time, an MU might be in the communication range of multiple entities that can contribute to learning and service delivery (contributors or participants). Before proceeding, we present a definition.
\begin{definition}[Global Aggregation]
A global aggregation at period $T$ refers to the required time to perform the cascaded FRPL and cache placement approaches until convergence, i.e., completing the involved tasks.
\end{definition}
\begin{figure*}
  \centering
  \includegraphics[width=11.00cm,height=5.10cm]{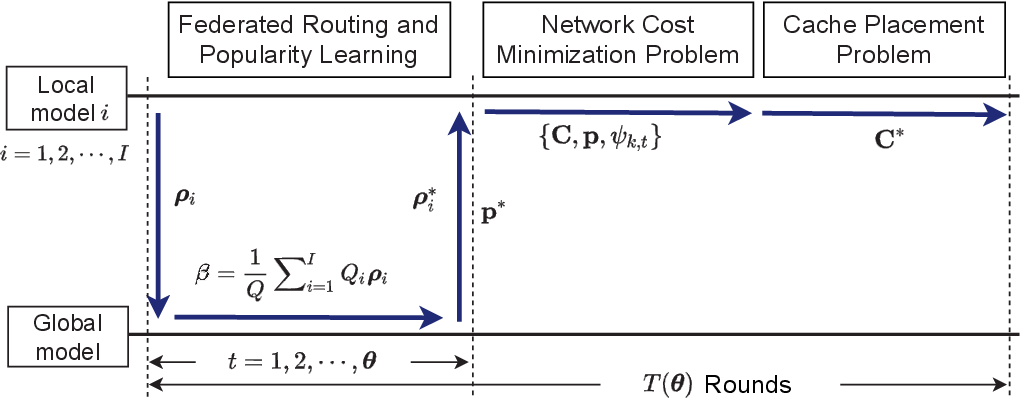}\\
  \caption{The learning procedure with ${I}$ participants.}
  \label{fig:01}
\end{figure*}
Throughout the paper, we use two time scales: (i) $t= \{1,2,\cdots, \boldsymbol{\theta}\}$. It refers to a learning round (period) in the FRPL method, where $\boldsymbol{\theta}$ is the required time until the global loss function converges; and (ii) $T(\boldsymbol{\theta}) = \{1,2,\cdots\}$, which corresponds every instance of the global aggregation. Thus, each global aggregation consists of several learning rounds. The proposed cache placement approach takes several iterations; however, at one time instant (learning period), both the routing strategy and cache placement remain fixed.

FL is the core of the proposed FRPL methodology. The MBS and SBSs
collaboratively learn a shared model while each entity retains its training data at its side. The FL model trained at the participant's side is the \textit{local model}. The MBS integrates the local models and generates the \textit{global model}, which improves the local model of each participant. \textbf{Fig. \ref{fig:01}} shows the workflow and details follow.

Consider $I$ participants labeled as $\{1,\cdots, {I}\}$.
We denote the input data of participant $i$, $i \in \{1,\cdots, {I}\}$ by $\mathbf{X}_{i} = [\mathbf{x}_{i,1}, \cdots, \mathbf{x}_{i,Q_{i}}]$, where $Q_{i}$ indicates the number of the samples. Thus $\mathbf{X}_{i}$ is the entry vector to train the local FL model. The output vector is $\mathbf{y}_{i} = [y_{i,1}, \cdots, y_{i,Q_{i}}]^{T}$. The vector $\boldsymbol{\rho}_{i}$ captures the model parameters of the local FL model. Under a linear model assumption, we have $\mathbf{y}_{i}= \mathbf{X}_{i} \boldsymbol{\rho}_{i}+\mathbf{n}$, where $\mathbf{n}$ is measurement noise typically approximated as Gaussian i.i.d. samples.
In the standard gradient descent (SGD) methods \cite{article201601}, \cite{mcmahan2017communication}, training and update procedures of the local FL model  aims at finding optimal parameters $\boldsymbol{\rho} _{i}$ that minimize the squared error cost function $f(\boldsymbol{\rho}_{i},\mathbf{X}_{i},  \mathbf{y}_{i})=||\mathbf{X}_{i}\boldsymbol{\rho}_{i}-\mathbf{y}_{i}||^{2}$.
We define a vector $\boldsymbol{\beta}$ to capture the parameters related to the global FL model. The update of the global model therefore is given by
$\boldsymbol{\beta}=\; \frac{1}{Q} \sum\nolimits_{i=1}^{{I}}\; Q_{i}  \boldsymbol{\rho}_{i}$. \textbf{Section \ref{sec:FedRout}} provides more details of the FRPL framework.
\subsection{Content Request Model}
\label{sec:Req}
We consider a content library with $M$ contents. The size of each file $f \in \{1,\cdots,M\}$ is $g_f$. The MUs request file $f$ randomly and independently with probability $p_{f} \in [0,1]$ so that $\mathbf{p} = [p_{1},\cdots,p_{M}]^{T}$ is the request probability vector. Traditionally, upon requesting a file within the time deadline $\tau$, an MU obtains it through random caching, local SBS caching, or MBS caching \cite{6708492}. Nevertheless, such naive caching mechanisms barely guarantee a high cache efficiency for the following reasons: (i) They largely neglect the limited storing capacity and the finite bandwidth; (ii) They often retransmit the data unnecessarily. Therefore, we decompose the cache domain of an MU's required contents into three categories:
\begin{itemize}
\item {\bf Local Caching:} The MU first checks the local SBS. If the required content is cached, then the MU obtains it directly from there within the time deadline $\tau$. Otherwise, the MU receives it from one of the following sources:
\item {\bf Intra-Cell Caching:} If the local SBS does not have the required content, it can fetch it from other SBSs in the intra-cell domain upon availability so that the MU is served within the deadline $\tau$.
\item {\bf Inter-Cell Caching:} If no SBS in the intra-cell domain has the required content, the local SBS fetches it via the backhaul link, from an external SBS deployed in the other overlapped cells or, in the worst case, from the MBS.
\end{itemize}
%
\subsection{Problem Statement}
\label{subsec:PrFor}
We focus on the following challenges: (i) When MUs migrate into one area, proper retrieving of several requested files over a backhaul while minimizing the cost becomes complicated; (ii) The MUs' preferences affect the cache efficiency of SBSs. Moreover, distributed caching might result in duplicate files in a small area; (iii) Joint learning and cache placement, also service provisioning, can yield long delays.

Let $\mathbf{C} \in \{0,1\}^{K \times M}$ be a binary caching strategy of the required contents in SBS, where $c^{}_{k,f}=1$ indicates the $k$th SBS stores the file $f$, and $c^{}_{k,f}=0$ indicates otherwise. Let $\sum \nolimits_{k=1}^{K} \sum\nolimits_{f=1}^{M} c^{}_{k,f} \cdot \alpha_{C_f}$ denote the SBS's \textit{aggregated cost} as a result of caching, where $\alpha_{C_f}$ is the cost of caching content $f$. The cost mainly depends on the file size $g_{f}$. The expected cost for retrieving content $f$ from SBS $k \in \{1,\cdots,K\}$ within a global aggregation period can be written as
\begin{align}
\label{equ:040101}
\begin{split}
J^{t}_{k,f} (\psi_{k,t}, p_{f}) = \psi_{k,t} \cdot p_{f} \cdot \left( c^{}_{k,f} \cdot \beta_{S_k}+
 (1-c^{}_{k,f}) \cdot d^{}_{k,f} \right),
\end{split}
\end{align}
where $\psi_{k,t}$ represents the pedestrian density at time slot $t$ at the site of SBS $k$. Besides, $\lambda_{f,t} = \psi_{k,t} \cdot p_{f}$ indicates the expected request density (i.e., the number of requests per time slot) of a specific file $f$ at time $t$. Finally, $d^{}_{k,f} = \prod\nolimits_{l\ne k}^{K}(1-c^{}_{l,f})(\beta_{M} + \beta_{M_k} + \beta_{S_k})$ corresponds to the worst case cost, i.e., when fetching file $f$ from the MBS backhaul. $\beta_{M}$ is a constant cost for retrieving $f$ via an MBS backhaul, $\beta_{M_k}$ denotes the cost of retrieving $f$ for SBS $k$ from MBS backhaul, and $\beta_{S_k}$ refers to the cost incurred by retrieving $f$ for MUs from the SBS $k$. Besides, $\lambda_{f,t} \cdot c^{}_{k,f} \cdot \beta_{S_k}$ is the expected cost for retrieving $f$ from SBS $k$, if available. Minimizing the aggregated cost $J^{t}_{k,f} (\psi_{k,t}, p_{f})$ in (\ref{equ:040101}) over the $K$ SBSs, i.e., $\mathop {\text{minimize} } \nolimits_{\{\mathbf{C}\}} \quad \sum\nolimits_{k=1}^{K} J^{t}_{k,f}(\psi_{k,t}, p_{f})$, yields the optimal routing strategy for each file. Such content routing strategy stipulates that retrieving file $f$ through the desired SBS $k^{*} = \argmin\nolimits_{} \quad \sum\nolimits_{k=1}^{K} J^{t}_{k,f}(\psi_{k,t}, p_{f})$ leads to a least aggregated cost.

The expected cost ${J_{k,f}^{t}(\psi^{}_{k,t}, p_{f})}$ in (\ref{equ:040101}) is valid for the costs for every file request, regardless of the file being cached or not; i.e., it can serve the general transmission cost function for small-cell networks. Besides, it is an upper bound for the transmission cost incurred for retrieving file $f$ from SBS $k$ in a global aggregation period. Note that, in our mobility-aware scenario, a multicase-based caching policy is complex to implement. Indeed, most of the papers in that direction model the requests by a  Poisson point process (PPP) across SBSs \cite{7370924}; however, such assumption is not valid in our case, as the mobility pattern, and, thus, the user-SBS association, is dynamic and unknown.

The optimal mobility-aware routing strategy and cache placement is equivalent to minimizing the network cost per global aggregation. Formally,
\begin{subequations}
\label{equ:06}
\begin{align}
& \mathop {\text{minimize}} \limits_{\{\mathbf{C} \}} \quad \sum\limits_{k=1}^{K}  \sum\limits_{f=1}^{M}  c^{}_{k,f} \cdot \alpha_{C_f}  + \sum\limits_{k=1}^{K}\sum\limits_{f=1}^{M} J^{t}_{k,f}(\psi_{k,t}, p_{f})  \\
& \;\;\; \text{s.t.} \sum\limits_{f=1 }^{M} c^{}_{k,f}g_{f} \le C_{k}^{S}, \forall \, k \in \{1,\cdots, {K}\},\, t= \{1,2,\cdots\},\\
& \quad\;\;\;\;\; c^{}_{k,f} \in \{0,1\}, \; \forall \; k , f ,
\end{align}
\end{subequations}
where Constraint (\ref{equ:06}b) means that the total size of cached files cannot exceed the cache capacity of SBS $k$.

Problem (\ref{equ:06}) is an integer programming problem that is NP-hard. Moreover, the objective function is not available since it involves \textit{unknown} popularity $p_{f}$ and pedestrian density $\psi_{k,t}$. In particular, there exists $2^{M+K}$ possible caching strategy matrices $\{\mathbf{C}\}$, implying an exponential growth in complexity as a function of $M$ and $K$. Therefore, it is essential to develop an efficient approach to solve the problem (\ref{equ:06}) while maintaining a low delay.
\section{Federated Routing and Popularity Learning}
\label{sec:FedRout}
Based on the OSS concept, we propose an FRPL approach to learn
the pedestrian- and request density while ensuring a fast model aggregation. It consists of three major steps:

{\textbf{Geographical Location Division:}} We divide the entire area uniformly into $K$ small parts. Each area includes an SBS and a set of connected MUs at time slot $t$, denoted by $\mathcal{U}_{k,t}, k \in \{1,\cdots, K\}$. Usually, MUs in $\mathcal{U}_{k,t}$ have the same network cell ID at this moment.

{\textbf{Dual-Task Execution:}} The SBS of each sub-area aims at learning the pedestrian- and files' request density by exploiting the location- and request information. Details follow.
\begin{itemize}
\item {\it TASK 1 (Pedestrian Density Prediction)}:
To predict the pedestrian density of a cell $k$, $k \in \mathcal{K}$, the corresponding SBS derives the following statistics for the set of MUs at $K$ areas: (i) The number of pedestrian clusters in the transition region of $\kappa$ neighboring cells that have $k$ as the predicted next cell \cite{7146028}; (ii) The number of pedestrians already transited to $k$; (iii) The number of pedestrians that probably leave cell $k$.\\
To search the $\kappa$ neighboring areas to find the candidates of pedestrian clusters, the SBS uses the $K$-means clustering algorithm \cite{bennett2000constrained} with the loss function given by $L_{1}({\boldsymbol{\rho}_{1,i}}):=\sum\nolimits_{j \in  \mathcal{U}_{i,t} }|| \mathbf{x}_{1,i,j}-f(\boldsymbol{\rho}_{1,i}, \mathbf{x}_{1,i,j})||^{2}$, where $f(\boldsymbol{\rho}_{1,i},\mathbf{x}_{1,i,j})$ is the centroid of all objects assigned to $x_{1,i,j}$'s class.\\
In words, the pedestrian clustering minimizes the sum of squared errors between data points and their respective centroids until reaching a stationary centroid.\\
To obtain the number of pedestrian clusters that are approaching the desired cell $k$, the SBS uses the following detection criterion:
\begin{align}
\label{equ:062705}
\frac{\sqrt{(x_{0}-x_{k}(t))^{2} + (y_{0}-y_{k}(t))^{2}}}{\sqrt{(x_{0}-x_{k}(t-1))^{2} + (y_{0}-y_{k}(t-1))^{2}}} <1,
\end{align}
where $(x_{0},y_{0})$ is the location of the SBS $k$, and $(x_{k}(t),y_{k}(t))$ is the final location of the centroid of cluster $k$ with $k=\{1,\cdots,\kappa\}$.
We use $N_{k,t}^{+}(j)$ to denote the statistical function of density associated with the pedestrians that might transit to cell $k$ from neighboring cell $j$. Besides, $N_{k,t}^{-}(j)$ is the statistical function of density associated with the pedestrians might leave cell $k$ to the neighboring cell $j$. Thus, the pedestrian density of cell $k$ from the neighboring cell $j$ yields $N_{k,j,t} = \text{max}(N_{k,t}^{+}(j) - N_{k,t}^{-}(j), 0)$. Moreover, $N_{k,0,t}^{}$ refers to the density function associated with the pedestrians already moved to cell $k$. Then the pedestrian density of the desired cell yields
\begin{align}
\label{equ:0627006}
\begin{split}
\psi^{*}_{k,t} &= N_{k,0,t}^{}+ \sum\limits_{i=1}^{\kappa} \text{max} \left( N_{k,t}^{+}(i) - N_{k,t}^{-}(i), 0\right) \\ &= N_{k,0,t}+N_{k,1,t} + \ldots + N_{k,\kappa,t} = \sum\limits_{i=0}^{\kappa} N_{k,i,t},
\end{split}
\end{align}
given that their respective centroids of clusters $1,\cdots,\kappa$ satisfy the detection criterion in (\ref{equ:062705}) accordingly. \textbf{Fig. \ref{fig:0002}} illustrates the procedure described above.
%
\begin{figure}[t]
  \centering
  \includegraphics[width=70mm,height=65mm]{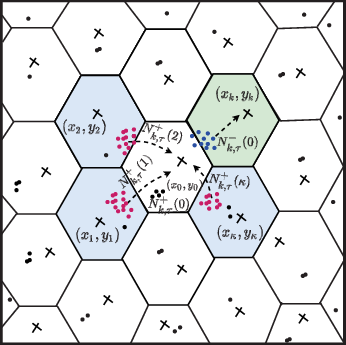}\\
  \caption{Prediction of $\psi_{k,t}^{*}$ in a dense network where moving pedestrians (purple dots) that might move to the desired cell (crosses) are clustered, whereas moving pedestrians (blue dots) that might leave the current cell are clustered at time slot $\tau$.}
  \label{fig:0002}
\end{figure}
%
\item {\it TASK 2 (Request Density Prediction)}:
The SBS first searches the $\kappa$ neighboring areas to find those MUs falling into $\mathcal{U}_{k,t}, k \in \{1,\cdots, K\}$, and to observe the demographic information about those associated users and their ratings. Naturally, the files with high ratings have more request density.

Using the files' rating together with the users' features, the SBS learns the request density of every file $f$, $f=\{1,\cdots, M\}$, by minimizing the least squared error between the estimated request density and the actual one. By exploring the linear regression model to predict $\lambda_{f,t}^{}$, the SBS uses the following loss function
\begin{align}
\label{equ:062306}
L_{2}({\boldsymbol{\rho}_{2,i}}):=\frac{1}{2} ||y_{2,i,j} - f(\boldsymbol{\rho}_{2,i}, \mathbf{x}_{2,i,j})||^{2} + \alpha ||\boldsymbol{\rho}_{2,i}||^{2},
\end{align}
where $\alpha$ is a regularization hyperparameter. The popularity of file $f$ is $p_{f} = \frac{\lambda_{f,t}}{\sum\nolimits_{f=1}^{M}\lambda_{f,t}}$.
Thus, the expected request density $\lambda_{f,t}^{*}$ is given by
\begin{align}
\label{equ:Lambda}
\begin{split}
\lambda_{f,t}^{*} &= \psi^{*}_{k,t} \cdot p_{f} \\ &= \frac{ \big(N_{k,0,t}+N_{k,1,t} + \ldots + N_{k,\kappa,t}\big) \cdot \lambda_{f,t}}{\sum\limits_{f=1}^{M}\lambda_{f,t}}.
\end{split}
\end{align}
Later in \textbf{Section \ref{Sec:CachePlacement}}, the SBS uses this metric for cache placement.
\end{itemize}
\textbf{Fast Model Aggregation:}
In standard SGD methods \cite{article201601}, \cite{mcmahan2017communication}, the unknown model is iteratively estimated by computing $\boldsymbol{\rho} _{q,i}$ at each epoch, evaluating a gradient associated to the loss function
\begin{align}
\label{equ:0302008}
L_{q}( \boldsymbol{\rho} _{q,i}) =
\begin{cases}
{\sum\nolimits_{j \in  \mathcal{U}_{i,t} }|| \mathbf{x}_{q,i,j}-f(\boldsymbol{\rho}_{q,i}, \mathbf{x}_{q,i,j})||^{2}}, & {q = 1},\\
{}\\
{\frac{1}{2} ||y_{q,i,j} - f(\boldsymbol{\rho}_{q,i}, \mathbf{x}_{q,i,j})||^{2} + \alpha ||\boldsymbol{\rho}_{q,i}||^{2}}, & {q = 2},
\end{cases}
\end{align}
with $q = \{1,2\}$ being the indices of prediction tasks.
Thus the training procedure of the FRPL finds the optimal parameters $\{\boldsymbol{\rho}_{q,1}, \cdots, \boldsymbol{\rho}_{q,K}\}$ that minimize the sum of squared  error cost  function, i.e., $L_{q}( \boldsymbol{\rho} _{q,i})$ in (\ref{equ:0302008}). For the linear regression FRPL in the request density prediction task, the model averaging constraint guarantees that $K$ SBSs and MBS share the same FL model after the convergence. \\
In the framework of FRPL, $K$ SBSs send the local models $\{\boldsymbol{\rho}_{q,1}, \cdots, \boldsymbol{\rho}_{q,K}\}$ to the MBS.\footnote{For each global aggregation, all participants update their parameters $\boldsymbol{\rho}_{q,1}, \cdots, \boldsymbol{\rho}_{q,K}$ to facilitate the highest running efficiency of FRPL \cite{yang2019scheduling}.}~At time slot $t$, the global model, denoted by $\boldsymbol{\beta}_{q,t}$, is given by\footnote{We denote the local and global model parameters with time-scale index $t$, to elaborate the procedure of aggregating the local and global FL models at each epoch.}
\begin{align}
\label{equ:051212}
\boldsymbol{\beta}_{q,t}=
\frac{1}{Q_{q}} \sum\limits_{i=1}^{K} \;{Q_{q,i} \cdot \boldsymbol{\rho}_{q,i,t}} &,  \;\;\quad q\in \{1,2\},
\end{align}
where $Q_{q,i}$ indicates the number of samples collected by the local model $i$ w.r.t. the task $q$. Besides, $Q_{q}= \sum\nolimits_{i=1}^{{K}} Q_{q,i}$ is the total number of samples collected by all local models w.r.t. the task $q$.
Note that $\sum\nolimits_{i=1}^{K} {Q_{q,i} \cdot \boldsymbol{\rho}_{q,i,t} }$ indicates the total number of training data points. The MBS then sends $\boldsymbol{\beta}_{q,t}$ back to the SBSs.\\
After receiving $\boldsymbol{\beta}_{q,t}$ from the MBS, the SBSs use the SGD method to update the local models $\{\boldsymbol{\rho}_{q,1}, \cdots, \boldsymbol{\rho}_{q,K}\}$. The update of local model $\boldsymbol{\rho}_{q,i,t}$ follows as \cite{article201601}
\begin{align}
\label{equ:051212}
\boldsymbol{\rho}_{q,i,t+1} =  \boldsymbol{\beta}_{q,t} - \frac{\eta   }{Q_{q,i}} \sum\limits_{j=1}^{Q_{q,i}} \nabla  f ( \boldsymbol{\beta}_{q,t}, \; \mathbf{x}_{q,i,j},\; {y}_{q,i,j})
\end{align}
w.r.t. the task $q$, where $\eta$ denotes the learning rate. Moreover, $\nabla  f(\boldsymbol{\beta}_{q,t},\mathbf{x}_{q,i,j},{y}_{q,i,j})$ corresponds to the gradient function of $f( \boldsymbol{\beta}_{q,t},\mathbf{x}_{q,i,j}, {y}_{q,i,j})$ w.r.t. $\boldsymbol{\beta}_{q,t}$. For simplicity, we hereby define $\small F(\boldsymbol{\beta}_{q,t},\mathbf{x}_{q,i,j},{y}_{q,i,j})=  \frac{1}{Q_{q}}\sum\nolimits_{i=1}^{K} \sum\nolimits_{j=1}^{Q_{q,i}} f(\boldsymbol{\beta}_{q,t} , \mathbf{x}_{q,i,j}, {y}_{q,i,j})$ as the squared error cost function of the global model.\\
After receiving the updated local models, the MBS also updates the global model $\boldsymbol{\beta}_{q,t}$ as \cite{9013160}
\begin{align}
\label{equ:032212}
\begin{split}
\boldsymbol{\beta}_{q,t+1} =
   \boldsymbol{\beta}_{q,t}  - \eta \Big(   \nabla  F(\boldsymbol{\beta}_{q,t}, \mathbf{x}_{q,i,j},  {y}_{q,i,j}) - \boldsymbol{\Theta}_{q}  \Big), \; q\in \{ 1, 2\},
\end{split}
\end{align}
where $\nabla F(\boldsymbol{\beta}_{q,t},\mathbf{x}_{q,i,j}, {y}_{q,i,j})$ is the gradient function of $F(\boldsymbol{\beta}_{q,t},\mathbf{x}_{q,i,j},{y}_{q,i,j})$ w.r.t. $\boldsymbol{\beta}_{q,t}$. Besides, $\boldsymbol{\Theta}_{q}$ is given by
\begin{align}
\small
\label{equ:040913}
\boldsymbol{\Theta}_{q} =  \nabla  F(\boldsymbol{\beta}_{q,t}, \mathbf{x}_{q,i,j},  {y}_{q,i,j}) - \frac{1}{Q_{q}} \sum\limits_{i=1}^{ {{K}}_{} } \; {Q_{q,i} \cdot \boldsymbol{\rho}_{q,i,t} }, \; q\in \{ 1, 2\}.
\end{align}
\textbf{Algorithm \ref{alg:Framwork}} summarizes the described stages.\\
%
\alglanguage{pseudocode}
\begin{algorithm}[htbp]
\caption{Federated Routing and Popularity Learning (FRPL)}
\label{alg:Framwork}
\begin{algorithmic}[1]
\Require
Initialize the points $T \leftarrow 0$, $t \leftarrow 0$, $\kappa \leftarrow 0$, and $\eta > 0$.
\Ensure The pedestrian density ${\psi}^{*}_{1:K_{}, \,t}$ and the expected request density $\lambda^{*}_{1:M, \,t}$.
\For{$T = 1,2, \cdots$}
\For{$t=1, \cdots, {\theta}$} \Comment{\textbf{Pedestrian Density Prediction}}
\State \multiline{
For each local FL model, the respective SBS searches the $\kappa$ neighboring areas to find the candidates of pedestrian clusters.}
\State \multiline{Calculate the gradient of the loss function $L_{1}( \boldsymbol{\rho} _{1,i,t})$.}
\State \multiline{Monitor the statistic of clusters and estimate the number of clusters by following the criterion in (\ref{equ:062705}).}
\State \multiline{Update the local FL model $\boldsymbol{\rho} _{1,i,t}$ and the centroids of $\kappa$ clusters.}
\State \multiline{Estimate the pedestrian density
$\psi^{*}_{k,t} \leftarrow N_{k,0,t}+N_{k,1,t} + \cdots + N_{k,\kappa,t}.$}
\EndFor
\For{$t=1,\,\cdots,\, \theta$} \Comment{\textbf{Request Density Prediction}}
\State \multiline{ Search the $\kappa$ neighboring areas and observe the statistical information in terms of users' demographic information as well as their ratings.}
\State \multiline{Pre-process data by merging the users' demographic information and rating information.}
\State \multiline{Learn the request density $\lambda_{f,t}$ of each file $f\in \{1,\cdots,M\}$, by means of minimizing the least squared error $L_{2}({\boldsymbol{\rho}_{2,i,t}})$ between the estimated and actual request density.}
\State \multiline{Update the local FL model $\boldsymbol\rho_{2,i,t}$.}
\State \multiline{Estimate the expected request density
$ \lambda^{*}_{f,t} \leftarrow \frac{ \big(N_{k,0,t}+N_{k,1,t} + \cdots + N_{k,\kappa,t}\big) \cdot \lambda_{f,t}}{\sum\nolimits_{f=1}^{M}\lambda_{f,t}}.$ }
\EndFor
\State \multiline{The distributed $K$ SBSs invoke the SGD method to update the local FL models $\{\boldsymbol{\rho}_{q,1,t}, \cdots, \boldsymbol{\rho}_{q,K_{},t}\}$, $q=\{1,2\}$, to MBS to aggregate the local FL models.}
\State {Update the global FL model $\boldsymbol{\beta}_{q,t}$ by following (\ref{equ:032212}).}
\State \multiline{Update the gradient function of global FL model by (\ref{equ:040913}).}
\EndFor
\end{algorithmic}
\end{algorithm}
%
Finally, we note that the dimension of the raw pedestrians' data (including geographic location, date, time, sojourn time) affects the outcome of $K$-means clustering algorithm \cite{bennett2000constrained}. We therefore state the following proposition.
\begin{proposition}
\label{pro:Character}
The expected request density $\lambda_{f,t}^{*}$ has the following properties:
\begin{itemize}
\item When the datasets utilized for user clustering in transition region of neighboring cells are insufficient (empty or very few data points), the associated $\lambda_{f,t}^{*}$ can be lower-bounded as
    \begin{align}
    \label{equ:0828013}
    \lambda_{f,t}^{*} \ge \frac{ N_{k,0,t} \cdot \lambda_{f,t}}{\sum\limits_{f=1}^{M}\lambda_{f,t}}.
    \end{align}
\item If the datasets' dimension is larger than $10$, the associated $\lambda_{f,t}^{*}$ can be upper-bounded as
    \begin{align}
    \label{equ:08280314}
    \lambda_{f,t}^{*} \le \frac{ \left(N_{k,0,t}+N_{k,1,t} + \ldots + N_{k,\kappa^{*},t}\right) \cdot \lambda_{f,t}}{\sum\limits_{f=1}^{M}\lambda_{f,t}},
    \end{align}
where $\kappa^{*}$ indicates the maximum number of non-empty clusters.
\end{itemize}
\end{proposition}
\begin{IEEEproof}
See Appendix A.
\end{IEEEproof}
\section{Cache Placement Policy}
\label{Sec:CachePlacement}
\begin{figure}[t]
  \centering
  \includegraphics[width=7.0cm,height=4.1cm]{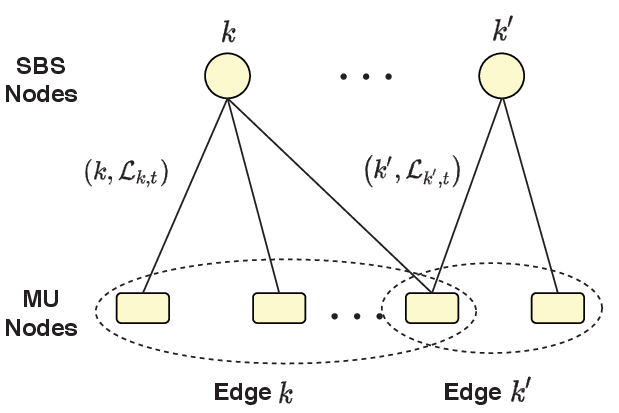}\\
  \caption{Selecting the serving edge at time slot $t$ with $(k, \mathcal{L}_{k,t})$ being the candidate edge.}
  \label{Fig:02}
\end{figure}

For each SBS $k \in\{1,\cdots,K\}$, let $\mathcal{L}_{k,t}$ denote the \textit{serving edge}, i.e., the set of MUs that it severs at time $t$ at both the intra- and inter-cell domains, as shown in \textbf{Fig. \ref{Fig:02}}. Obviously, the $\mathcal{L}_{k,t}$ is time-variant and depends on the mobility of users.\\
For each global aggregation, every SBS $k$ attempts to serve as many MUs as possible across its service edge to reduce the aggregated cost of retrieving $M$ contents across $K$ SBSs, i.e., $\sum\nolimits_{k=1}^{K}\sum\nolimits_{f=1}^{M} J^{t}_{k,f}(\psi_{k,t}, p_{f})$ in (\ref{equ:06}). Thus, for simplicity, we assume that at each global aggregation, the maximum number of MUs that SBS $k$ serves in its service edge equals to the maximum pedestrian density $\psi^{}_{k,t}$ that FRPL method predicts.\footnote{By \textbf{Proposition 1}, at time $t$, the pedestrian density of each cell $k$ can be upper bounded by $N_{k,0,t}+N_{k,1,t} + \cdots + N_{k,\kappa^{*},t}$, where $\kappa^{*}$ corresponds to the maximum number of non-empty clusters; That is, the maximum number of MUs that SBS $k$ can serve is equivalent to the upper bound of the pedestrian density of this cell.}
\subsection{The Algorithm}
After learning $\psi^{*}_{k,t}$ and $p^{}_{f}$ using the FRPL method, each SBS $k \in\{1,\cdots,K\}$ knows also $\mathcal{L}^{*}_{k,t}$. Then the optimal cache placement problem is equivalent to minimizing the network cost function $\mathcal{D}(\{c^{}_{k,f}\})$ given by
\begin{align}
\label{equ:1101016}
\begin{split}
& \mathcal{D}(\{c^{}_{k,f}\})
= \sum\limits_{k=1}^{K}\sum\limits_{f=1}^{M} c^{}_{k,f} \cdot \alpha_{C_f}+\\ & \sum\limits_{k=1}^{K} \sum\limits_{f=1}^{M} \psi_{k,t}^{*} \cdot p_{f} \cdot \Big( c^{}_{k,f} \cdot \beta_{S_k}+(1-c^{}_{k,f}) \cdot d^{}_{k,f}\Big).
 \end{split}
\end{align}
Thus the optimization problem follows as
\begin{subequations}
\label{equ:0406014}
\begin{align}
 &  \mathop {\text{minimize} }\limits_{\{c^{}_{k,f}\}}\quad\quad\quad\quad {\mathcal{D}}(\{c^{}_{k,f}\})   \\
&  \,\,\, \text{s.t.}  \sum\limits_{f=1 }^{M} c^{}_{k,f}g_{f} \le C_{k}^{S}, \; \forall\; k\in\{1,\cdots,K\}, \, t= \{1,2,\cdots\},\\
& \;\;\;\;\;\,\, \; c^{}_{k,f} \in \{0,1\}, \;\quad \forall \; k , f.
\end{align}
\end{subequations}
Problem (\ref{equ:0406014}) is an NP-hard integer programming. Given cache placement policy $\{c^{}_{k,f}\}$, the content retrieval policy is determined based on our content transmission protocol introduced in \textbf{Section \ref{subsec:PrFor}}. To solve the problem with low complexity, we develop \textbf{Algorithm \ref{alg:Framwork3}}, which greedily places the files to a cache and provides an approximate solution. \\
Let $I_{k}$ be the total size of files that are already cached at SBS $k$ during each iteration of our algorithm. Besides, $\mathcal{L}_{k,f}$ denotes the set of pairs $\{k,f\}$, where SBS $k$ does not store file $f$ yet despite having the sufficient cache capacity. At iteration ${\Theta}$, the algorithm conducts the cache placement by picking the pair $\{k^{*},f^{*}\} \in \mathcal{L}^{\Theta}_{k,f}$ with the lowest $\mathcal{D}(\{c^{}_{k,f}\})$, provided that such cost is lower than the one in the previous iteration. Note that if the algorithm starts with the pair $\{k,f^{*}\}$ that results in the lowest cost $\mathcal{D}(\{c^{}_{k,f^{*}}\})$, then the relative network cost caused by SBS $k$ at each iteration remains fixed. Formally,
\begin{align}
\label{equ:0406015}
\{k^{*},f^{*}\} =\argmin\limits_{\{k^{},f^{}\} \in \mathcal{L}^{{\Theta}}_{k,f}}
{\mathcal{D}}(\{c^{}_{k,f}\}).
\end{align}
Then, if $I^{{\Theta}}_{k^{*}} < C^{S}_{k^{*}}$, the algorithm updates $I^{{\Theta+1}}_{k^{*}} = I^{{\Theta}}_{k^{*}} + c^{}_{k^{*},f^{*}} \cdot g_{f^{*}}$ and $\mathcal{L}^{{\Theta+1}}_{k,f} = \mathcal{L}^{\Theta}_{k,f}\backslash \{k^{*},f^{*}\}$; Otherwise, when $I^{\Theta}_{k^{*}} = C^{S}_{k^{*}}$ implies that the cache of SBS $k^{*}$ is full, the algorithm excludes all pairs $\{k^{*},f\}$ from $\mathcal{L}^{\Theta}_{k,f}$, and stops the cache placement for SBS $k^{*}$. The algorithm terminates when all the caches are full. \textbf{Algorithm \ref{alg:Framwork3}} summarizes the described greedy cache placement.
\alglanguage{pseudocode}
\begin{algorithm}[!h]
\caption{Cache Placement Policy}
\label{alg:Framwork3}
\begin{algorithmic}[1]
\Require Initialize the occupied cache size $I_{k}$ with zero. Let $K_{ite}$ be the required number of iterations.
\Ensure The near-optimal cache placement $\{c^{}_{k,f}\}$.
\For{{$\Theta$} = 1, 2, $\cdots$, $K_{ite}$}
\State \multiline{ Pick the pair $\{k^{*},f^{*}\} \in \mathcal{L}_{k,f}$ with the lowest network cost $\mathcal{D}(\{c^{}_{k,f}\})$ according to (\ref{equ:0406015}).}
\If{$I^{\Theta}_{k^{*}} < C^{S}_{k^{*}}$}
\State {Let $I^{\Theta+1}_{k^{*}} \leftarrow I^{\Theta}_{k^{*}} + c^{}_{k^{*},f^{*}} \cdot g_{f^{*}}$.}
\State {Let $\mathcal{L}^{\Theta+1}_{k,f} \leftarrow \mathcal{L}^{\Theta}_{k,f}\backslash \{k^{*},f^{*}\}$.}
\Else {\;Exclude all the pairs $\{k^{*},f^{}\}$ $\forall f$ from $\mathcal{L}^{\Theta}_{k^{},f}$.}
\EndIf
\EndFor
\end{algorithmic}
\end{algorithm}
\subsection{Discussion}
Let $B_{(k)_{u},f}$ be a binary variable that indicates whether SBS $k$ caches a file $f$ requested by MU $u \in \mathcal{L}^{*}_{k,t}$. We observe that \textit{the approximation ratio} of our proposed cache policy is proportional to the number of MUs that an SBS serves at each global aggregation \cite{Kellerer2004knapsack} when neglecting the normalized cost for retrieving files from MBS.\footnote{Numerical analyses show that the normalized cost parameter for retrieving files from MBS backhaul has only a limited influence on the network cost minimization.}
To support this conclusion, we present the following results.
\begin{proposition}
For each global aggregation period, the normalized cost caused by caching at SBSs is equal to the aggregated costs for caching request contents over the serving edges. Then we have
\begin{align}
\label{equ:1202026}
\begin{split}
\sum\limits_{\{k,f\} \in \mathcal{L}_{k,f}^{}} c_{k,f}^{} \cdot \alpha_{C_f} =
\sum\limits_{\{k,f\} \in \mathcal{L}_{k,f}^{}} {\rm{min}}\Big(1,\sum\limits_{u \in \mathcal{L}^{*}_{k,t}} B_{(k)_{u},f}\Big) \cdot \alpha_{C_f},
\end{split}
\end{align}
where $\mathcal{L}^{*}_{k,t}$ is the estimated service edge of SBS $k \in \{1,\cdots,K\}$ at time $t$.
\end{proposition}
\begin{IEEEproof}
See Appendix B.
\end{IEEEproof}
\begin{theorem}
When the normalized cost for retrieving files from MBS backhaul is neglected, i.e., for $d^{}_{k,f}$ = 0,
our proposed cache placement policy in {\textbf{Algorithm \ref{alg:Framwork3}}} is a polynomial-time $L$-approximation algorithm with $L$ being the maximum number of MUs that an SBS serves at its service edge at each time.
\end{theorem}
\begin{IEEEproof}
See Appendix C.
\end{IEEEproof}
\subsection{Complexity Analysis}
Based on the optimization criterion given by (\ref{equ:0406015}), the computational complexity of cache placement using {\textbf{Algorithm \ref{alg:Framwork3}}} at SBS $k$ is \textit{polynomial time} $\mathcal {O}\left( n \cdot \left\|\mathcal{L}_{k,f}\right\|_{0} \right)$, with $n$ being the required number of iterations to solve problem (\ref{equ:0406015}). As $\mathcal{L}_{k,f}$ is the set of all possible (SBS, file) pairs that impose the lowest ${\mathcal{D}}(\{c^{}_{k,f}\})$ to fill the cache of SBS $k$, the total computational complexity of the cache placement policy yields $\mathcal {O}\left( \sum\nolimits_{k=1,\cdots,K} n \cdot \left\|\mathcal{L}_{k,f}\right\|_{0} \right)$. Hence the proposed policy has a linear complexity in the number of (SBS, file) pair.
\begin{table}[!t]
\linespread{1.1} 
\captionsetup{font={footnotesize,sc}, format=plain, singlelinecheck=off, labelsep=newline}
 \captionsetup{justification=centering}
\tocaption{SYSTEM PARAMETER SETUPS}
  \centering
  \footnotesize
  \begin{tabular}{|l|l|}
    \hline
    \textbf{Parameters} & \textbf{Value}  \\
     \hline
      \hline
      Global aggregation periods (rounds) $T$ &  $3.4 \times 10^{3}$  \\ \hline
     The volume of content library  $M$ & $3952$ \\  \hline
     The total number of MUs $U$ & 6040   \\ \hline
       The learning rate $\eta$ & $3\times 10^{-3}$  \\  \hline
       The hyperparameter for regularization, i.e., $\alpha$ & [1, 10]  \\  \hline
       The size of each content $f$, i.e., $g_{f}$ & 1MB \\ \hline
       Cache size of SBS $k$, i.e., $C^{S}_{k}$  & [50MB:500MB]   \\  \hline
      Probability of randomly selecting $m$ contents, i.e., $\epsilon $ & [0.01, 0.1, 0.8] \\  \hline
      Power cost of caching content $f$, i.e., $\alpha_{C_f}$ & 1.5 mW  \\  \hline
      Power cost of retrieving $f$ via MBS, i.e., $\beta_{M}$ & 13 mW \\ \hline
      Power cost of retrieving $f$ for SBS $k$ from MBS, i.e., $\beta_{M_k}$ & 370 mW   \\ \hline
      Power cost of retrieving $f$ via SBS $k$, i.e., $\beta_{S_k}$ & 180 mW \\ \hline
  \end{tabular}
\label{Tb:Setting}
\end{table}
\section{Simulation Results and Analysis}
\label{sec:simulation}
%
\subsection{Datasets and System Parameter Setups}
We use a real-world dataset, namely, \textit{MovieLens 1M} \cite{10.5555Maxwell}, to evaluate our proposed learning and caching strategies. Similar to \cite{7775114}, we assume that the movie rating process in the datasets is a streaming request. The \textit{MovieLens 1M} datasets consist of (i) the users' demographic information given in the format of \textless user ID*, age (in 7 categories), gender (in 2 categories), occupation (in 21 categories), Zip-code\textgreater; (ii) the rating information given in the format of \textless user ID*, movie ID, rating (in 5 categories), timestamp\textgreater.\\
To predict the request density, in the experiments, we merge the data of users' demographic- and rating information based on the label of the user ID*. Against this join-type dataset, we pick the attributes of \textless movie ID, user ID*, their gender, age, Zip-code, and occupation\textgreater ~as the context dimensions. \textbf{Table \ref{Tb:Setting}} gathers the most important parameters of our experiments.
\subsection{Performance Evaluation}
We evaluate the average caching efficiency (the average ratio of cache hits within one global aggregation period compared to the total requests) of our proposals in comparison with the following cache placement approaches:
%
\begin{figure}[t]
\centering
\includegraphics[width=75mm,height=55mm]{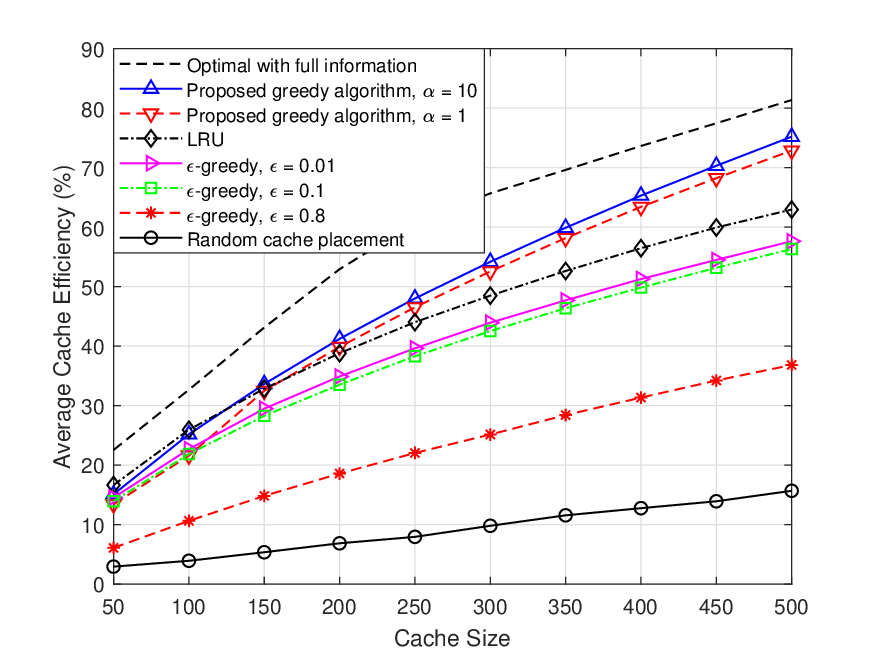}\\
\caption{The average cache efficiency (CE) of different cache placement approaches.}
\label{Fig:04}
\end{figure}
\begin{figure}[t]
\centering
\includegraphics[width=75mm,height=55mm]{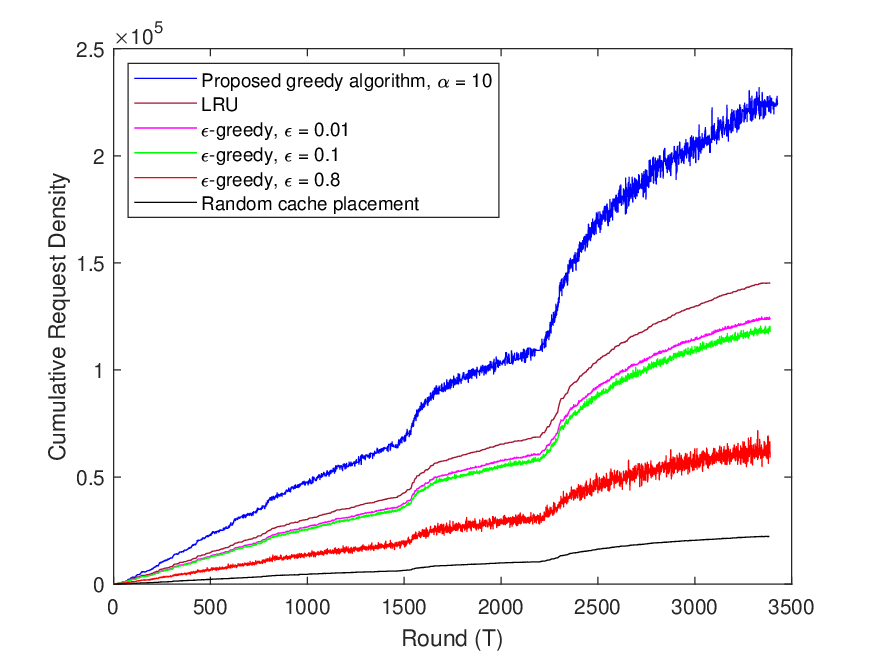}\\
\caption{Cumulative request density of different cache placement approaches.}
\label{Fig:05}
\end{figure}
\begin{itemize}
\item \textbf{Optimal with Full Information:} This scheme has full information about the users' demands; thus, it has the potential of providing the best caching performance;
\item {\textbf{$\boldsymbol{\epsilon}$-Greedy}:} The policy selects a set of $m$ files uniformly at random with probability $\epsilon \ll 1$, while with probability $1-\epsilon $, it selects the $m$ files with the highest estimated popularity so far;
\item \textbf{Least-Recently-Used (LRU):} It fetches the requested files from the potential contributor (i.e., MBS backhaul) and caches it when a cache miss event occurs. If an item exceeds the cache capacity of SBS, it removes the file in the cache that has been least recently used;
\item \textbf{Random:} This scheme selects a random set of files to cache in each time slot.
\end{itemize}
In \textbf{Fig. \ref{Fig:04}}, we compare the performance of different caching approaches using the same dataset, so that the file set remains constant. The figure shows that our proposed algorithm achieves a gain significantly higher than that of the random- and $\epsilon$-greedy approaches under different $\epsilon$. The gain grows as the cache size increases. The reason is that the proposed algorithm greedily places the cache to SBSs after learning the expected request densities of files by \textbf{Algorithm \ref{alg:Framwork}}. \\
Besides, our proposed approach exhibits superior performance compared to the LRU approach for large storage space. Also, by comparison between the proposed and optimal approaches, we deduce that the proposed algorithm provides a near-optimal solution while pursuing a minimum network cost.
\begin{figure*}[t]
\center
    \begin{minipage}[t]{0.43\linewidth}
     \begin{center}
    \includegraphics[width=75mm,height=55mm]{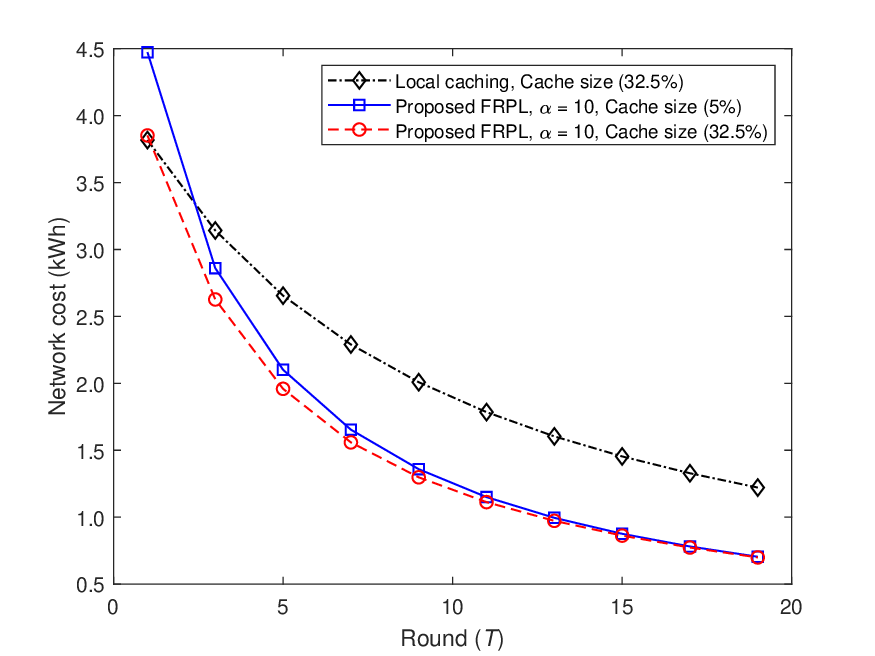}
    \end{center}
    \centerline{\;\; \qquad \small{(a)} }
    \end{minipage}
    \begin{minipage}[t]{0.43\linewidth}
       \begin{center}
    \includegraphics[width=75mm,height=55mm]{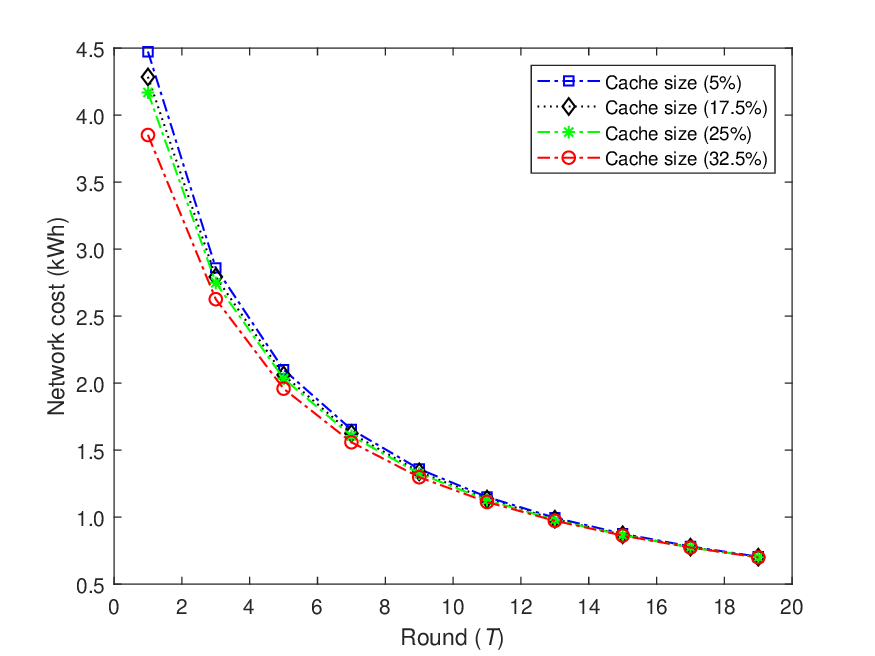}
    \end{center}
    \centerline{\;\; \qquad \small{(b)} }
    \end{minipage}
    \caption{Network cost impacted by (a) distinct cache placement methods; and (b) the proposed FRPL approach using distinct average cache sizes ($\%$ of file library) with cache content update duration = 4 (Hours).} \label{Fig:06}
\end{figure*}

\textbf{Fig. \ref{Fig:05}} shows the cumulative request density of the cached files up to time slot $T$ with $C_{k}^{S}=200$ MB. The {\it cumulative request density} is the sum of request densities of desired files at each round. The figure shows that the proposed algorithm achieves larger cumulative request density than $\epsilon$-greedy and random caching methods. In particular, at time slot $T = 2500$, the cumulative request density achieved by the proposed greedy algorithm is $1.65$, $1.87$, $1.93$, $3.76$, and $10.31$ times the cumulative request density achieved by LRU, $\epsilon$-greedy (for $\epsilon=0.01,0.1,0.8$), and random placement, respectively.

In \textbf{Fig. \ref{Fig:06}}, we evaluate the network cost of the proposed FRPL- and local caching methods. It can be seen from \textbf{Fig. \ref{Fig:06}(a)} that the cost of our cache placement solution is less than that of local caching. The figure also shows that, compared to the local caching, the FRPL approach requires fewer iterations to arrive at a given level of network cost. This is because the FRPL approach accurately learns the expected request density from the context of observed files' rating and users' features. \textbf{Fig. \ref{Fig:06}(b)} depicts the network cost of the proposed FRPL approach ($\alpha$ = 10) for distinct cache sizes. We assume that the caching entity (i.e., the SBS) updates its cache content every 4-hours. \textbf{Fig. \ref{Fig:06}(b)} then shows that expanding the cache from 5$\%$ to 32.5$\%$ reduces the network cost of the FRPL approach. The reason is that enlarging the cache size expands the solution space of problem (\ref{equ:0406014}).
%
\begin{figure*}[t]
\center
    \begin{minipage}[t]{0.43\linewidth}
     \begin{center}
    \includegraphics[width=75mm,height=55mm]{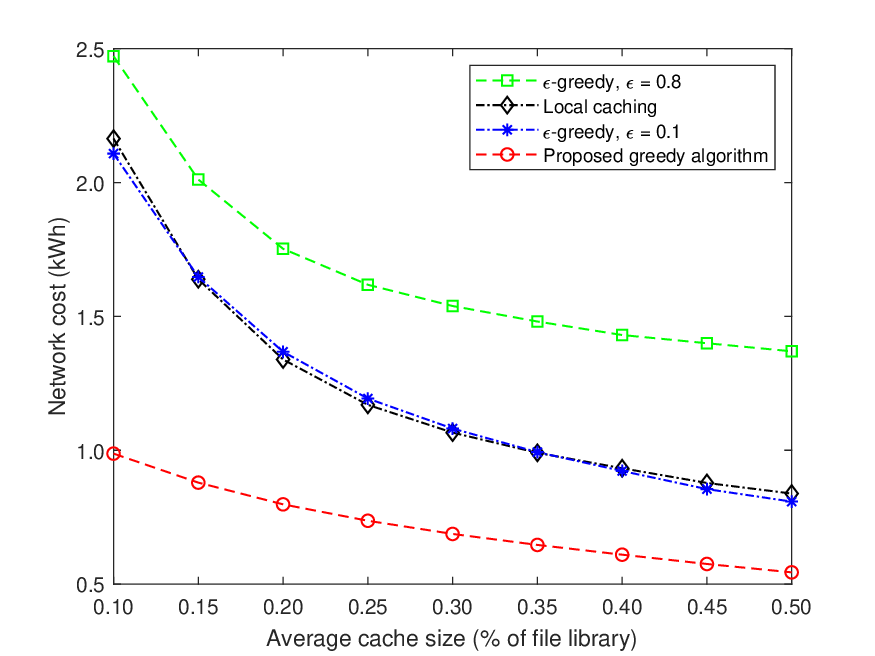}
    \end{center}
    \centerline{\;\; \qquad \small{(a)}}
    \end{minipage}
    \begin{minipage}[t]{0.43\linewidth}
       \begin{center}
    \includegraphics[width=75mm,height=55mm]{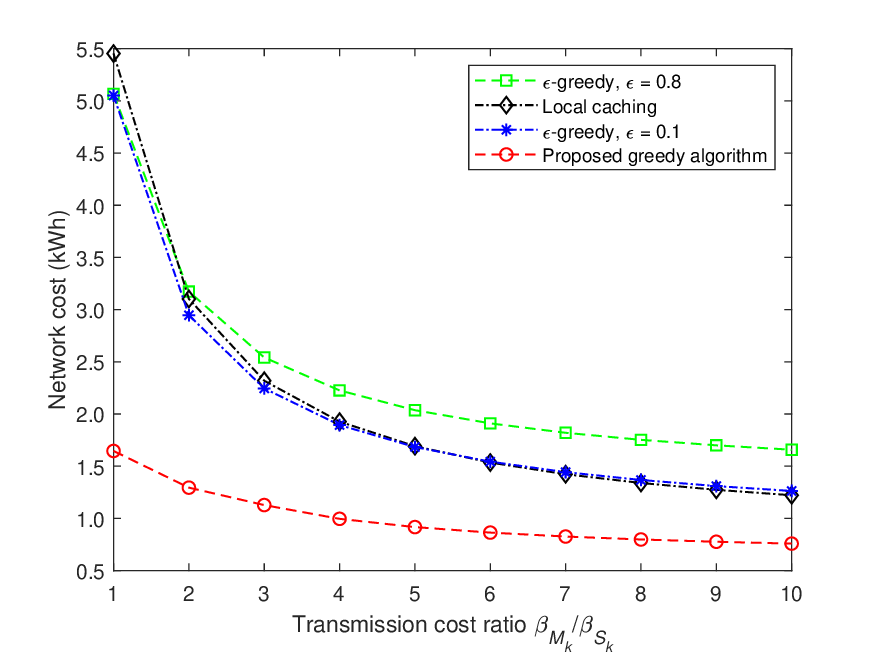}
    \end{center}
    \centerline{\;\; \qquad \small{(b)} }
    \end{minipage}
    \caption{Network cost of different placement methods with (a)
    distinct average cache sizes ($\%$ of file library); and (b) different transmission cost ratios $\beta_{M_{k}}/\beta_{S_{k}}$.}
    \label{Fig:07}
\end{figure*}
\begin{figure*}[t]
\center
    \begin{minipage}[t]{0.43\linewidth}
     \begin{center}
    \includegraphics[width=75mm,height=55mm]{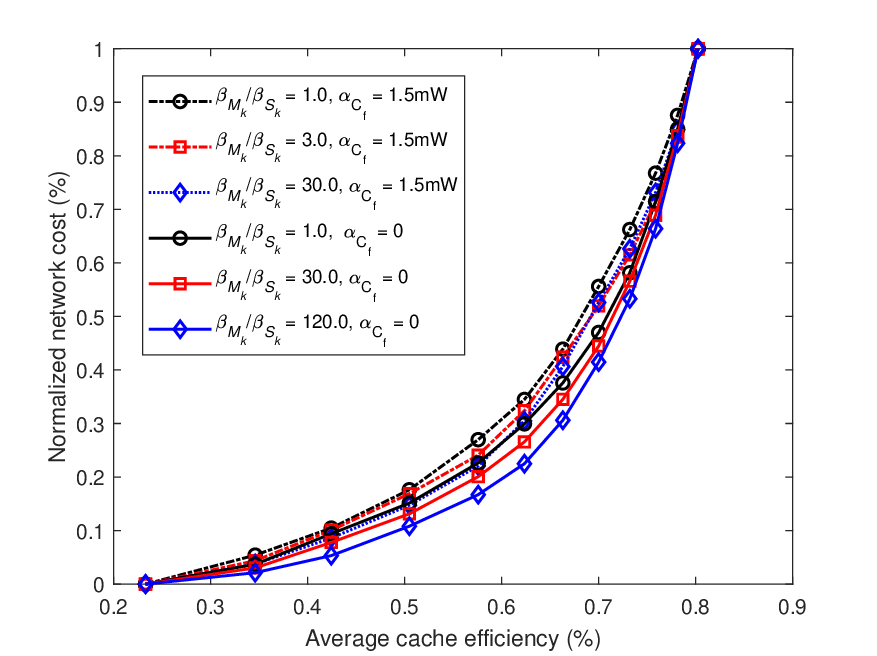}
    \end{center}
    \centerline{\;\; \qquad \small{(a)} }
    \end{minipage}
    \begin{minipage}[t]{0.43\linewidth}
       \begin{center}
    \includegraphics[width=75mm,height=55mm]{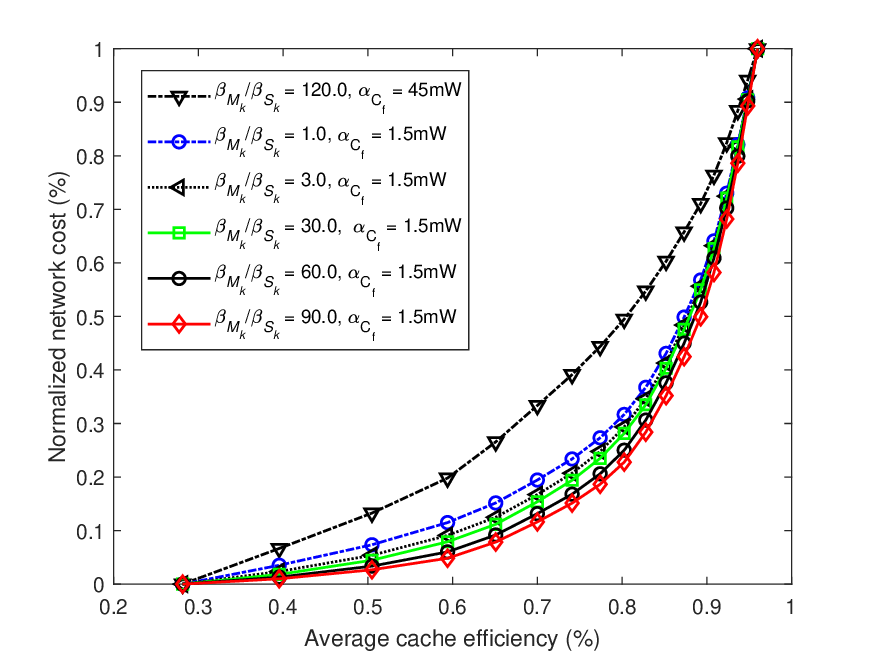}
    \end{center}
    \centerline{\;\; \qquad \small{(b)} }
    \end{minipage}
    \begin{minipage}[t]{0.43\linewidth}
       \begin{center}
    \includegraphics[width=75mm,height=55mm]{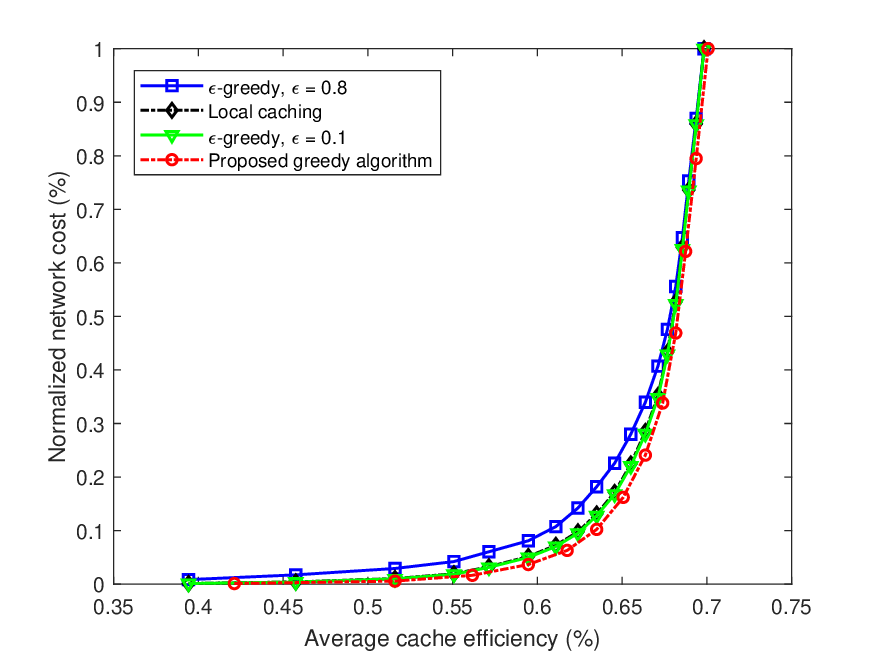}
    \end{center}
    \centerline{\;\; \qquad \small{(c)} }
    \end{minipage}
    \caption{The normalized network cost versus the average cache efficiency with (a) distinct transmission cost ratios $\beta_{M_{k}}/\beta_{S_{k}}$ and the costs of caching content $f$, the cache update duration is 15 hours; (b) different transmission cost ratios $\beta_{M_{k}}/\beta_{S_{k}}$ and the costs of caching content $f$, the cache update duration is 20 hours; and (c) different cache placement methods, the cache update duration is 20 hours.}
    \label{Fig:08}
\end{figure*}

\textbf{Fig. \ref{Fig:07}(a)} and \textbf{Fig. \ref{Fig:07}(b)} show the network cost of different placement methods as a function of the cache size ($\%$ of file library) and the transmission cost ratio $\beta_{M_{k}}/\beta_{S_{k}}$, respectively. \textbf{Fig. \ref{Fig:07}(a)} shows that the proposed algorithm reduces the network cost compared to $\epsilon$-greedy ($\epsilon=0.1,0.8$), and local caching approaches for cache sizes varying from 10$\%$ to 50$\%$. The result in \textbf{Fig. \ref{Fig:07}(a)} is consistent with that in \textbf{Fig. \ref{Fig:06}(a)}. Besides, \textbf{Fig. \ref{Fig:07}(b)} shows that the transmission cost ratio $\beta_{M_{k}}/\beta_{S_{k}}$ dramatically influences the cost performance of different methods. In particular, when increasing $\beta_{M_{k}}/\beta_{S_{k}}$ from 1 to 5, the cost of different placement approaches falls abruptly. Indeed, the parameter of normalized cost of file retrieval from the SBSs affects the performance more than that from the MBS over a backhaul.

In \textbf{Fig. \ref{Fig:08}(a)-\ref{Fig:08}(c)}, we evaluate the normalized network cost versus the average caching efficiency as a function of the transmission cost ratio $\beta_{M_{k}} / \beta_{S_{k}}$ and the cost of caching content $f$ for different placement approaches. We utilize movie rating data collected from \textit{MovieLens 1M} \cite{10.5555Maxwell} associated with distinct cache content update intervals. To model the user demand process, we synthesize two datasets with cache content update intervals equal to 15 hours (\textbf{Fig. \ref{Fig:08}(a)}) and 20 hours (\textbf{Fig. \ref{Fig:08}(b)} and \textbf{Fig. \ref{Fig:08}(c)}). \textbf{Fig. \ref{Fig:08}(a)} shows that: (i) with the decreasing cost of caching content $f$, i.e., $\alpha_{C_f}$, the proposed placement approach reduces the normalized network cost; (ii) with the increasing transmission cost ratio $\beta_{M_{k}}/\beta_{S_{k}}$ and for a given $\alpha_{C_f}$, our approach improves the cost performance. Besides, \textbf{Fig. \ref{Fig:08}(b)} shows that when facing ultra-high request traffic, our approach reduces the normalized network cost. Finally, \textbf{Fig. \ref{Fig:08}(c)} reveals that the proposed placement approach reduces the network cost compared to the $\epsilon$-greedy ($\epsilon=0.1,0.8$) and local caching methods. The reason is that our proposal accurately predicts the expected request density of each file particularly for high-dimensional datasets.
\section{Concluding Remarks and Future Directions}
\label{sec:Conclusion}
We developed mobility-aware routing and caching strategies for resource-constrained small-cell networks in an FL framework. To jointly optimize the routing and cache placement, we first formulated the NP-hard network cost minimization problem. To address the complexity issue, we first proposed an approach by which the SBSs learn the pedestrian density and request density. Afterward, based on the predictions, a greedy algorithm optimizes the cache placement and minimizes the network cost. Numerical results established that our proposed method yields a near-optimal solution.

Potential future works include the extension of our proposal in the following directions:\\
(i) Integrating energy-efficient real-time routing architectures using machine learning methods to reduce the routing overhead \cite{tavli2011energy} and discovery latency \cite{imielinski1996mobile} in mobile edge networks.\\
(ii) Allowing for time-variant popularity (i.e., dynamic content library), which renders learning, prediction, and energy-efficient network performance challenging.\\
(iii) Finally, it is essential to investigate a proper offloading of the learning tasks to potential participants, under constraints such as spectrum scarcity, transmission latency, and the like.
\begin{appendices}
\section{Proof of Proposition 1.}
\label{sec:Proof1}
\textit{$K$-means clustering} performs poor on datasets with $n \ge 10$ dimensional real space $\mathbb{R}^{n}$. As a result, the $K$-means clustering often converges with at least one or more clusters with either empty or very few data points. In this case, to facilitate fast model aggregation, we discard the empty clusters or summarize very few data points when performing the pedestrian density prediction. That results in $\kappa^{*}$ non-empty clusters. Cluster filtration is essential, as a modest value of $\kappa^{*}$ enables solutions that summarize the underlying data better \cite{bennett2000constrained}.\\
Thus, when the datasets for user clustering in transition regions of neighboring cells are scarce, the estimated pedestrian density $\psi^{*}_{k,t}$ reaches trough of $N_{k,0,t}$. That results in a lower bound $\lambda_{f,t}^{*} \ge \frac{ N_{k,0,t} \cdot \lambda_{f,t}}{\sum\nolimits_{f=1}^{M}\lambda_{f,t}}$. In contrast, when the number of dimensions of datasets is larger than $10$,
the estimated pedestrian density $\psi^{*}_{k,t}$ reaches at a peak value of $\left(N_{k,0,t}+N_{k,1,t} + \ldots + N_{k,\kappa^{*},t}\right)$, resulting in an upper bound
$\lambda_{f,t}^{*} \le \frac{ \left(N_{k,0,t}+N_{k,1,t} + \ldots + N_{k,\kappa^{*},t}\right) \cdot \lambda_{f,t}}{\sum\nolimits_{f=1}^{M}\lambda_{f,t}}$, which completes the proof. $\hfill\blacksquare$
\section{Proof of Proposition 2.}
\label{sec:Proof3}
The normalized caching cost at $K$ SBSs is given by
\begin{align}
\label{equ:1202027}
\sum \limits_{k=1}^{K} \sum\limits_{f=1}^{M} c^{}_{k,f} \cdot \alpha_{C_f}.
\end{align}
Based on the definition of $\mathcal{L}_{k,f}$, $(\ref{equ:1202027})$ can be rewritten as
\begin{align}
\label{equ:1202028}
\sum \limits_{\{k,f\} \in \mathcal{L}_{k,f}} c^{}_{k,f} \cdot \alpha_{C_f}.
\end{align}
Then, by the definition of $B_{(k)_{u},f}$, the aggregated cost for caching the request contents at the serving edge equals
\begin{align}
\label{equ:1202029}
\sum \limits_{\{k,f\} \in \mathcal{L}_{k,f}} {\rm{min}}\Big(1,\sum\limits_{u \in \mathcal{L}^{*}_{k,t}} B_{(k)_{u},f}\Big) \cdot \alpha_{C_f}.
\end{align}
When no user requests content $f$ that is in the cache of SBS $k$, the involved normalized cost $c^{}_{k,f} \cdot \alpha_{C_f}$ and also the the aggregated cost at the serving edge, i.e., ${\rm{min}}\big(1,\sum\nolimits_{u \in \mathcal{L}^{*}_{k,t}} B_{(k)_{u},f}\big) \cdot \alpha_{C_f}$, become zero.
If that event occurs, one concludes that the normalized cost $\sum \nolimits_{\{k,f\} \in \mathcal{L}_{k,f} } c^{}_{k,f} \cdot \alpha_{C_f}$ in (\ref{equ:1202028}) equals the aggregated cost of MUs that fall into the edge $\mathcal{L}^{*}_{k,t}$, i.e., $\sum \nolimits_{\{k,f\} \in \mathcal{L}_{k,f}} {\rm{min}}\big(1,\sum\nolimits_{u \in \mathcal{L}^{*}_{k,t}} B_{(k)_{u},f}\big) \cdot \alpha_{C_f}$ in (\ref{equ:1202029}).\\
When some MUs request a content $f$ that is not in the cache, the proposed placement policy selects a unique $(k^{*},f)$ pair from $\mathcal{L}_{k,f}$ associated with the lowest ${\mathcal{D}}(\{c^{}_{k^{*},f}\})$. This means that $c^{}_{k^{*},f} = 1$. In this case, we have $c^{}_{k^{*},f} \cdot \alpha_{C_f} = \alpha_{C_f}$. Meanwhile, ${\rm{min}}\big(1,\sum\nolimits_{u \in \mathcal{L}^{*}_{k^{*},t}} B_{(k^{*})_{u},f}\big)=1$. That is because a requested content $f$ that falls into the set ${\mathcal{L}^{*}_{k^{*},t}}$ will be cached at SBS deployed at either the intra-cell domain or the inter-cell domain, so that ${\rm{min}}\big(1,\sum\nolimits_{u \in \mathcal{L}^{*}_{k^{*},t}} B_{(k^{*})_{u},f}\big) \cdot \alpha_{C_f} = \alpha_{C_f} $. Therefore the equality of the normalized cost in (\ref{equ:1202028}), i.e., traversing all the candidate $\{k,f\}$ pairs of the set $\mathcal{L}_{k,f}$, equaling to the aggregated cost in (\ref{equ:1202029}) holds, when an event of requesting an uncached content $f$ occurs.\\
Based on the discussion above we have
\begin{align}
\label{equ:1202030}
\begin{split}
& \sum \limits_{\{k,f\} \in \mathcal{L}_{k,f} } c^{}_{k,f} \cdot \alpha_{C_f} =
\\ & \sum \limits_{\{k,f\} \in \mathcal{L}_{k,f} } {\rm{min}}\Big(1,\sum\limits_{u \in \mathcal{L}^{*}_{k,t}} B_{(k)_{u},f}\Big) \cdot \alpha_{C_f}.
\end{split}
\end{align}
This completes the proof of Proposition 2. $\hfill\blacksquare$
%

\section{Proof of Theorem 1.}
\label{sec:Proof4}
The network cost minimization problem is given by
\begin{align}
\label{equ:1202032}
\begin{split}
 \mathcal{D}(\{c^{}_{k,f}\})
= & \sum\limits_{k=1}^{K}  \sum\limits_{f=1}^{M}  c^{}_{k,f} \cdot \alpha_{C_f}  + \sum\limits_{k=1}^{K}\sum\limits_{f=1}^{M}\psi_{k,t}^{*} \cdot p_{f} \cdot  c^{}_{k,f}  \cdot \beta_{S_k} \\
& + \sum\limits_{k=1}^{K}\sum\limits_{f=1}^{M} \psi_{k,t}^{*} \cdot p_{f} \cdot (1-c^{}_{k,f}) \cdot d^{}_{k,f}.
 \end{split}
\end{align}
Let $\mathcal{A}$ and $\mathcal{B}$ be two distinct subsets of $\mathcal{A}_{t}$, i.e., $\mathcal{A} \subseteq \mathcal{A}_{t}$, $\mathcal{B} \subseteq \mathcal{A}_{t}$, and $\mathcal{A} \neq \mathcal{B}$. When keeping in the summation only a subset of the terms and all the terms are positive \cite{7370924}, i.e., $c^{}_{k,f} \ge 0$, $\alpha_{C_f} \ge 0$, $\lambda_{f,t}^{*} = \psi_{k,t}^{*} \cdot p_{f} \ge 0$, $\beta_{S_k} \ge 0$, $(1-c^{}_{k,f}) \ge 0$, and $d^{}_{k,f} \ge 0$, the following inequalities hold, i.e., $\sum\nolimits_{k=1}^{K}  \sum\nolimits_{f=1}^{M}  c^{}_{k,f} \cdot \alpha_{C_f} \ge \sum\nolimits_{\{k,f\} \in \mathcal{A}} c^{}_{k,f} \cdot \alpha_{C_f}$, $\sum\nolimits_{k=1}^{K}\sum\nolimits_{f=1}^{M} \lambda_{f,t}^{*} \cdot c^{}_{k,f}  \cdot \beta_{S_k} \ge \sum\nolimits_{\{k,f\} \in \mathcal{A}} \lambda_{f,t}^{*} \cdot c^{}_{k,f}  \cdot \beta_{S_k}$, and $\sum\nolimits_{k=1}^{K}\sum\nolimits_{f=1}^{M} \lambda_{f,t}^{*} \cdot (1-c^{}_{k,f}) \cdot d^{}_{k,f} \ge \sum\nolimits_{\{k,f\} \in \mathcal{B}} \lambda_{f,t}^{*} \cdot (1-c^{}_{k,f}) \cdot d^{}_{k,f}$.
Thus, the network cost $\mathcal{D}(\{c^{}_{k,f}\})$ in (\ref{equ:1202032}) can be rewritten by
\begin{align}
\label{equ:1204033}
\begin{split}
 \mathcal{D}(\{c^{}_{k,f}\})
&= \sum\limits_{k=1}^{K}  \sum\limits_{f=1}^{M}  c^{}_{k,f} \cdot \alpha_{C_f}  + \sum\limits_{k=1}^{K}\sum\limits_{f=1}^{M} \lambda_{f,t}^{*} \cdot c^{}_{k,f}  \cdot \beta_{S_k} \\
& \quad + \sum\limits_{k=1}^{K}\sum\limits_{f=1}^{M} \lambda_{f,t}^{*} \cdot (1-c^{}_{k,f}) \cdot d^{}_{k,f}\\
& \geq \sum\limits_{\{k,f\} \in \mathcal{A}} c^{}_{k,f} \cdot \alpha_{C_f}+
\sum\limits_{\{k,f\} \in \mathcal{A}} \lambda_{f,t}^{*} \cdot c^{}_{k,f}  \cdot \beta_{S_k}  \\
& \quad +   \sum\limits_{\{k,f\} \in \mathcal{B}}  \lambda_{f,t}^{*} \cdot (1-c^{}_{k,f}) \cdot d^{}_{k,f}\\
& \overset{(a)} {\geq} \sum\limits_{\{k,f\} \in \mathcal{A}} c^{}_{k,f} \cdot \alpha_{C_f}+
\sum\limits_{\{k,f\} \in \mathcal{A}} \lambda_{f,t}^{*} \cdot c^{}_{k,f}  \cdot \beta_{S_k},
\end{split}
\end{align}
where inequality $(a)$ holds if one neglects the normalized cost for retrieving files from MBS, i.e., $d^{}_{k,f}=0$.

Based on {\textbf{Proposition 2}}, we arrive at the following equivalent transformations against the normalized cost for caching contents in (\ref{equ:1202027}) and the normalized cost for retrieving contents in (\ref{equ:1202032}) in the form of
\begin{align}
\label{equ:1203034}
\begin{split}
& \sum\limits_{\{k,f\} \in \mathcal{A}} c^{}_{k,f} \cdot \alpha_{C_f} =  \sum \limits_{\{k,f\} \in \mathcal{A}}   \Bigg\{{\rm{min}}\Big(1,\sum\limits_{u \in \mathcal{L}^{*}_{k,t}} B_{(k)_{u},f}\Big)  \cdot \alpha_{C_f} \Bigg\},
\end{split}
\end{align}
\begin{align}
\label{equ:1203035}
\begin{split}
 & \sum\limits_{\{k,f\} \in \mathcal{A}}  \lambda_{f,t}^{*} \cdot c^{}_{k,f}   \cdot \beta_{S_k}  \\ & =\sum\limits_{\{k,f\} \in \mathcal{A}} \Bigg\{\lambda_{f,t}^{*} \cdot {\rm{min}}\Big(1,\sum\limits_{u \in \mathcal{L}^{*}_{k,t}} B_{(k)_{u},f}\Big)  \cdot \beta_{S_k} \Bigg\}.
\end{split}
\end{align}
Substituting (\ref{equ:1203034}) and (\ref{equ:1203035}) back into (\ref{equ:1204033}) yields (\ref{equ:1202033}), shown at the top of this page.

Based on the facts that ${\rm min} \big(1, \sum\nolimits_{i=1,\cdots,I}x_{i} \big) \ge \frac{1}{I} \sum\nolimits_{i=1,\cdots,I} {\rm min} \left(1, x_{i}\right) $, we therefore can arrive at that ${\rm{min}}\big(1,\sum\nolimits_{u \in \mathcal{L}^{*}_{k,t}} B_{(k)_{u},f}\big) \ge \frac{1}{\left\|\mathcal{L}^{*}_{k,t}\right\|_{0}} \sum\nolimits_{u \in \mathcal{L}^{*}_{k,t}} {\rm{min}} \big(1, B_{(k)_{u},f}\big)$. Substituting this lower bound approximation into (\ref{equ:1202033}) results in (\ref{equ:1202034}), shown at the top of this page.
%
\begin{figure*}[t]
\begin{align*}
\small
\label{equ:1202033} \tag{26}
\begin{split}
 \mathcal{D}\left(\{c^{}_{k,f}\}\right) & = \sum\limits_{k=1}^{K}  \sum\limits_{f=1}^{M}  c^{}_{k,f} \cdot \alpha_{C_f}  + \sum\limits_{k=1}^{K}\sum\limits_{f=1}^{M} \lambda_{f,t}^{*} \cdot c^{}_{k,f}  \cdot \beta_{S_k}
 \quad + \sum\limits_{k=1}^{K}\sum\limits_{f=1}^{M} \lambda_{f,t}^{*} \cdot (1-c^{}_{k,f}) \cdot d^{}_{k,f}\\
& \geq \sum\limits_{\{k,f\} \in \mathcal{A}} c^{}_{k,f} \cdot \alpha_{C_f}+
\sum\limits_{\{k,f\} \in \mathcal{A}} \lambda_{f,t}^{*} \cdot c^{}_{k,f}  \cdot \beta_{S_k}
 +   \sum\limits_{\{k,f\} \in \mathcal{B}} \lambda_{f,t}^{*} \cdot (1-c^{}_{k,f}) \cdot d^{}_{k,f}\\
& \geq \sum \limits_{\{k,f\} \in \mathcal{A}} \left\{{\rm{min}}\Big(1,\sum\limits_{u \in \mathcal{L}^{*}_{k,t}} B_{(k)_{u},f}\Big) \cdot \alpha_{C_f} \right\} +  \sum\limits_{\{k,f\} \in \mathcal{A}} \Bigg\{\lambda_{f,t}^{*} \cdot {\rm{min}}\Big(1,\sum\limits_{u \in \mathcal{L}^{*}_{k,t}} B_{(k)_{u},f}\Big)  \cdot \beta_{S_k} \Bigg\},
\end{split}
\end{align*}
\end{figure*}
%
\begin{figure*}[t]
\small
\begin{align*} \tag{27}
\label{equ:1202034}
\begin{split}
 \mathcal{D}\left(\{c^{}_{k,f}\}\right)
& \geq \sum \limits_{\{k,f\} \in \mathcal{A}} \left\{{\rm{min}}\Big(1,\sum\limits_{u \in \mathcal{L}^{*}_{k,t}} B_{(k)_{u},f}\Big) \cdot \alpha_{C_f} \right\} +  \sum\limits_{\{k,f\} \in \mathcal{A}} \Bigg\{\lambda_{f,t}^{*} \cdot {\rm{min}}\Big(1,\sum\limits_{u \in \mathcal{L}^{*}_{k,t}} B_{(k)_{u},f}\Big)  \cdot \beta_{S_k} \Bigg\} \\
& \geq \sum \limits_ {\{k,f\} \in \mathcal{A}}  \Bigg\{ \frac{1}{||\mathcal{L}^{*}_{k,t}||_{0}} \sum\limits_{u \in \mathcal{L}^{*}_{k,t}} {\rm{min}} \Big(1, B_{(k)_{u},f}\Big) \cdot \alpha_{C_f} \Bigg\} +  \sum\limits_{\{k,f\} \in \mathcal{A}}  \Bigg\{\lambda_{f,t}^{*}  \cdot \beta_{S_k} \cdot \frac{1}{||\mathcal{L}^{*}_{k,t}||_{0}} \cdot \sum\limits_{u \in \mathcal{L}^{*}_{k,t}}{\rm{min}}\Big(1, B_{(k)_{u},f}\Big) \Bigg\} \\
& = \frac{1}{||\mathcal{L}^{*}_{k,t}||_{0}} \sum\limits_{\{k,f\} \in \mathcal{A}} \Bigg\{ \sum\limits_{u \in \mathcal{L}^{*}_{k,t}} {\rm{min}} \Big(1, B_{(k)_{u},f}\Big) \cdot \alpha_{C_f} \Bigg\} +  \frac{1}{||\mathcal{L}^{*}_{k,t}||_{0}}  \sum\limits_{\{k,f\} \in \mathcal{A}} \Bigg\{\lambda_{f,t}^{*}  \cdot \beta_{S_k} \cdot  \sum\limits_{u \in \mathcal{L}^{*}_{k,t}}{\rm{min}}\Big(1, B_{(k)_{u},f}\Big) \Bigg\},
\end{split}
\end{align*}
\hrulefill
\end{figure*}

Next, we let $c^{*}_{k,f}$ be the optimal cache placement solution to the optimization problem (\ref{equ:0406015}).
In addition, define $B^{*}_{(k)_{u},f}$ as an optimal binary variable deciding whether file $f$ requested by MU $u \in \mathcal{L}^{*}_{k,t}$ is selected to be cached.
Thus we have $\sum \nolimits_{\{k,f\} \in \mathcal{A}} \Big\{ \sum\nolimits_{u \in \mathcal{L}^{*}_{k,t}} {\rm{min}} \big(1, B_{(k)_{u},f}\big) \cdot  \alpha_{C_f} \Big\} \ge \sum \nolimits_{\{k,f\} \in \mathcal{A}} \Big\{ \sum\nolimits_{u \in \mathcal{L}^{*}_{k,t}} {\rm{min}} \big(1, B^{*}_{(k)_{u},f}\big) \cdot \alpha_{C_f} \Big\}$, and $\sum\nolimits_{\{k,f\} \in \mathcal{A}} \Big\{\lambda_{f,t}^{*} \cdot   \sum\nolimits_{u \in \mathcal{L}^{*}_{k,t}}{\rm{min}}\big(1, B_{(k)_{u},f}\big) \cdot \beta_{S_k} \Big\} \ge \sum\nolimits_{\{k,f\} \in \mathcal{A}} \Big\{\lambda_{f,t}^{*} \cdot \sum\nolimits_{u \in \mathcal{L}^{*}_{k,t}}{\rm{min}}\big(1, B^{*}_{(k)_{u},f}\big) \cdot  \beta_{S_k}  \Big\}$.
Besides, by following the properties in {\textbf{Proposition 2}}, we have that $\sum\nolimits_{\{k,f\} \in \mathcal{L}_{k,f}^{K_{ite}}} c_{k,f}^{*} = \sum\nolimits_{\{k,f\} \in \mathcal{L}_{k,f}^{K_{ite}}} {\rm{min}}\big(1,\sum\nolimits_{u \in \mathcal{L}^{*}_{k,t}} B^{*}_{(k)_{u},f}\big)$ with $K_{ite}$ being the required number of iterations of {\textbf{Algorithm \ref{alg:Framwork3}}}. In addition, let $\mathcal{D}(\{c^{}_{k,f}\} | d^{}_{k,f} = 0)$ be the network cost of small-cell networks given that $d^{}_{k,f}$ = 0. Eventually, we can arrive at
\begin{align*}
\label{equ:1202035} \tag{28}
\begin{split}
& \mathcal{D}\left(\{c^{}_{k,f}\} | d^{}_{k,f} = 0\right) \\
& \geq \frac{1}{||\mathcal{L}^{*}_{k,t}||_{0}} \sum \limits_{\{k,f\} \in \mathcal{A}} \Bigg\{ \sum\limits_{u \in \mathcal{L}^{*}_{k,t}} {\rm{min}} \Big(1, B^{*}_{(k)_{u},f}\Big) \cdot \alpha_{C_f} \Bigg\} \\
& +  \frac{1}{||\mathcal{L}^{*}_{k,t}||_{0}}  \sum\limits_{\{k,f\} \in \mathcal{A}} \Bigg\{\lambda_{f,t}^{*} \cdot \beta_{S_k} \cdot \sum\limits_{u \in \mathcal{L}^{*}_{k,t}}{\rm{min}}\Big(1, B^{*}_{(k)_{u},f}\Big) \Bigg\} \\
& = \frac{1}{||\mathcal{L}^{*}_{k,t}||_{0}} \mathcal{D}\left(\{c^{*}_{k,f}\} | d^{}_{k,f} = 0 \right).
\end{split}
\end{align*}
By following (\ref{equ:1202035}), we conclude that $\mathcal{D}\big(\{c^{}_{k,f}\} | d^{}_{k,f} = 0 \big) \ge  \frac{1}{L} \mathcal{D}\big(\{c^{*}_{k,f}\} | d^{}_{k,f} = 0 \big)$ where $L = ||\mathcal{L}^{*}_{k,t}||_{0}$.
Based on the above observations, we thereby deduce that when $d^{}_{k,f}$ = 0, our proposed cache placement policy in {\textbf{Algorithm \ref{alg:Framwork3}}} indeed is of a polynomial-time $L$-approximation algorithm.
This completes the proof of Theorem 1. $\hfill\blacksquare$
\end{appendices}
\bibliographystyle{IEEEtran}
\bibliography{bibfile}
\end{document}